\documentclass[epj]{svjour}
\usepackage{graphicx}
\usepackage{amsmath}

\journalname{Eur. Phys. J. A}
\usepackage{amssymb}
\newcommand{\be}{\begin{eqnarray}}
\newcommand{\ee}{\end{eqnarray}}

\usepackage{color}

\begin{document}

\title{Deuteron transverse densities in holographic QCD}
\author{Chandan Mondal$^{1,2}$ \and Dipankar Chakrabarti$^2$ \and Xingbo Zhao$^1$}

\institute{$^1$Institute of Modern Physics, Chinese Academy of Sciences, Lanzhou 730000, China\\
$^2$Department of Physics, Indian Institute of Technology Kanpur, Kanpur-208016, India}

\date{Received: date / Revised version: date}

\abstract{We investigate the transverse charge density in the longitudinally as well as transversely polarized deuteron using the recent empirical description of the deuteron electromagnetic form factors in the framework of holographic QCD. The predictions of the holographic QCD are compared with the results of a standard phenomenological parameterization. In addition, we evaluate GPDs and the gravitational form factors for the deuteron. The longitudinal momentum densities are also investigated in the transverse plane.  
} 
\authorrunning{C. Mondal \and D. Chakrabarti\and X. Zhao}
\titlerunning{Deuteron transverse densities in holographic QCD}
\maketitle
\vskip0.2in
\noindent

\section{ Introduction}
The electromagnetic and gravitational form factors are among the most fundamental quantities containing information about the internal structure of the hadron. The Fourier transformation of these form factors provides information about spatial
distributions such as the charge and magnetization distribution and the longitudinal momentum distribution inside the hadron. During the last decade, many experimental and theoretical investigations have been focused on deuteron in hadronic physics. For detailed reviews, see Refs.~\cite{Abbott:2000ak,Garcon:2001sz,Kohl:2008zz,Holt:2012gg}. 
Since the deuteron has a spin of unity, there are three form factors, charge $G_C$, magnetic $G_M$,
and quadrupole $G_Q$. These form factors completely describe cross sections \cite{Buchanan:1965zz,Elias:1969mi,Benaksas:1964zz,Benaksas:1966zz,Platchkov:1989ch,Galster:1971kv,Cramer:1986kv,Simon:1981br,Abbott:1998sp,Berard:1974ev,Bosted:1989hy,Auffret:1985tg,Arnold:1986jda} and tensor polarizations \cite{Schulze:1984ms,Garcon:1993vm,The:1991eg,Abbott:2000fg} or  tensor analyzing powers \cite{Dmitriev:1985us,Gilman:1990vg,FerroLuzzi:1996dg,Bouwhuis:1998jj,Nikolenko:2003zq,Zhang:2011zu} which have been probed through the electron-deuteron elastic scattering.
There are many theoretical approaches such as chiral effective and phenomenological approaches, perturbative QCD, potential and quark
models, holographic QCD models etc. which have been extensively used to investigate the deuteron form factors \cite{Abbott:2000ak,Garcon:2001sz,Kohl:2008zz,Holt:2012gg,Brodsky:1983vf,Lev:1999me,Arnold:1979cg1,Arnold:1979cg2,De Forest:1966dn,Logunov:1963yc,Donnelly:1975ze,Friar:1975zza,Mathiot:1989vw,Schiavilla:1990ug,Allen:2000xy,Blankenbecler:1965gx,Buck:1979ff,Carbonell:1998rj,Karmanov:1991fv,Kolling:2012cs,Kolling:2012cs1,Buchmann:1989ua,Ito:1989nv,Dong:2008mt,Gutsche1,Gutsche2,Lyubo} whereas the generalized parton distributions (GPDs) for deuteron has been introduced in \cite{Berger:2001zb}. GPDs for the deuteron have been studied based on a phenomenological effective Lagrangian approach in \cite{Dong:2013rk}.
The gravitational form factors of spin-1 particles have been evaluated using a holographic model of QCD in \cite{AC4} where via sum rules, the connection of gravitational form factors to GPDs has also been established. Recently,
Gutsche $et.~al.$ \cite{Gutsche1,Gutsche2,Lyubo} have shown that the electromagnetic form factors for the deuteron in the soft-wall AdS/QCD model are well in agreement with the experimental  data and the form factors display correct $1/Q^{10}$ power scaling for large $Q^2$ which is consistent with the quark counting rules. Thus, it is very interesting to investigate the transverse charge densities, gravitational form factors and longitudinal momentum densities, GPDs for the deuteron in the framework of soft-wall AdS/QCD. 

Recently, tremendous interests have grown in AdS/ QCD formalism which emerges as one of the most promising tool to investigate the structure of hadrons. Though AdS/QCD gives only the semiclassical approximation of QCD, so far this formalism has been successfully applied to describe various hadronic properties e.g., hadron mass spectrum, parton distribution functions (PDFs), generalized parton distribution functions (GPDs), meson and nucleon form factors, transverse densities, structure functions etc.\cite{AC4,BT00,deTeramond:2010ge,BT01,BT011,katz,SS,AC,ads1,ads101,ModelII,ModelII1,ads2,BT1,Grigoryan:2007my,BT2,deTeramond:2005su,vega,vega01,CM,CM2,CM3,HSS,Mondal,Ma_baryon,reso,Chakrabarti:2016lod,BT_reso,BT_new1,BT_new2,BT_new3,deTeramond:2013it,Sufian:2016hwn}. In nuclear physics, there are also some applications of holographic QCD such as holographic nuclear matter~\cite{Kim:2007xi}, $\rho$ meson condensation at finite isospin chemical 
potential~\cite{Aharony:2007uu}, nuclear matter to strange matter transition~\cite{Kim:2009ey}, baryon matter at finite 
temperature and baryon number density~\cite{Bergman:2007wp}, 
cold nuclear matter~\cite{Rozali:2007rx}, heavy atomic nuclei \cite{Hashimoto:2008jq}, mean-field theory for baryon many-body systems~\cite{Harada:2011aa}. Many other interesting works have been done in nuclear physics using AdS/QCD formalism (see Refs. \cite{Kim:2011ey,Aoki:2012th,Kim:2012ey,Pahlavani:2014dma} for detailed reviews). 

The charge densities in the transverse plane are defined as the two dimensional Fourier transforms of the charge form factors. The transverse densities are also intimately related to the GPDs with zero skewness \cite{burk,burk01}. Since the form factor involves initial and final states with different momenta, three dimensional Fourier transforms cannot be interpreted as densities. However, the transverse densities defined at fixed light-front time have a proper density interpretation \cite{miller09,miller10,venkat}. The charge densities in the transverse impact parameter plane for the nucleon have been discussed extensively in various phenomenological models \cite{ModelII,ModelII1,CM3,miller07,vande,selyugin,CM4,CM401,neetika,Mondal:2016xpk,weiss}, whereas the densities in the transverse coordinate plane have been studied in \cite{hwang,Kumar:2014coa,Mondal:2016xsm}. In \cite{Liang:2014hca}, the deuteron transverse charge density has been evaluated using different parameterizations of the charge form factor and it has been shown that the different parameterizations provide different densities in particular at the center of the impact parameter ($b=0$). The deuteron transverse charge density has been studied using the phenomenological Lagrangian approach in \cite{Liang:2013pqa}. Using the parameterization of the deuteron form factor data \cite{Abbott:2000ak}, the transverse charge density in the deuteron has been mapped out in \cite{Carlson:2008zc} where the authors considered both the longitudinally and transversely polarized deuteron. In a similar fashion of charge density in the transverse plane, one can map out the longitudinal momentum distribution within a hadron by taking the two dimensional Fourier transformation of the gravitational form factor \cite{abidin08,selyugin}. 
The longitudinal momentum distributions within nucleons based on the phenomenological parameterization of GPDs and a similar distributions for spin-1 objects using theoretical results from the AdS/QCD correspondence have been calculated in \cite{abidin08}. A nice comparative investigation of charge and momentum density distributions has been done in \cite{selyugin} where the authors used a different $t$-dependence of GPDs from the Ref.\cite{abidin08} with same quarks distributions. Using a light front quark-diquark model in AdS/QCD, the longitudinal momentum densities have been evaluated for both the unpolarized and the transversely polarized nucleons in~\cite{CMA,CMA1}. A comparative study of the longitudinal momentum distribution for the nucleon in two different soft-wall AdS/QCD has been presented in \cite{Mondal}. In the present paper, using the recent empirical description of the deuteron electromagnetic form factors in the framework of soft-wall AdS/QCD, we investigate the transverse charge densities within a deuteron. We consider both the longitudinally and transversely polarized deuteron. In addition, we evaluate the GPDs and gravitational form factors for the deuteron which are further used to study the longitudinal momentum densities in the transverse plane.

This paper is organized as follows. A brief description of the electromagnetic form factors of the deuteron in the soft-wall AdS/QCD model has been given in Sec.\ref{DFFs}. The charge densities in the transverse plane for both unpolarized and transversely polarized deuterons have been analyzed in Sec.\ref{densities}. In Sec.\ref{DGPDs}, we present the deuteron GPDs and the results of gravitational form factors and the longitudinal momentum densities have been evaluated in Sec.\ref{Ldensity}. Finally, we provide a brief summary in Sec.\ref{concl}.
\section{Deuteron form factors}\label{DFFs}
The matrix element which describes the interaction of the deuteron 
with the electromagnetic field relates three form factors as  
\be\label{M_inv} 
&&M_{\rm inv}^\mu(p,p') =  - 
\Bigl( G_1(Q^2) \epsilon^\ast(p') \cdot \epsilon(p) \nonumber\\
&&-  \frac{G_3(Q^2)}{2M_d^2} \, \epsilon^\ast(p') \cdot q \, 
\epsilon(p) \cdot q \Bigr) \, (p+p')^\mu \nonumber\\
&-& G_2(Q^2) \, 
\Bigl( \epsilon^\mu(p) \, \epsilon^{\ast}(p') \cdot q 
- \epsilon^{\ast\mu}(p') \, \epsilon(p) \cdot q \Bigr) \, ,
\ee 
where $G_{1,2,3}$ are the form factors; $\epsilon$($\epsilon^\ast$) and
$p(p^\prime)$ are the polarization and four-momentum of the initial(final) deuteron, and $q=p^\prime - p$ is the momentum transfer. The electromagnetic form factors
$G_{1,2,3}$ of the deuteron
are again related to the charge $G_C$, quadrupole $G_Q$ and magnetic $G_M$
form factors by
\be
G_C &=& G_1+\frac{2}{3}\eta G_Q\,,\\ 
G_M &=& G_2 \,,            \\
G_Q &=& G_1-G_2+(1+\eta)G_3\,,  
\ee
with $\eta=\frac{Q^2}{4M_d^2}$ and normalization 
$
G_C(0)=1,
G_Q(0)=M_d^2{\cal Q}_d=25.83 ,
$
and $G_M(0)=\frac{M_d}{M_N}\mu_d=1.714,$
where $M_d$ and $M_N$ 
are the deuteron and nucleon masses, and 
${\cal Q}_d = 7.3424$ GeV$^{-2}$ and $\mu_d = 0.8574$ 
are the quadrupole and magnetic moments of the deuteron.
\begin{figure*}[htbp]
\begin{minipage}[c]{0.98\textwidth}
{(a)}\includegraphics[width=5.4cm,clip]{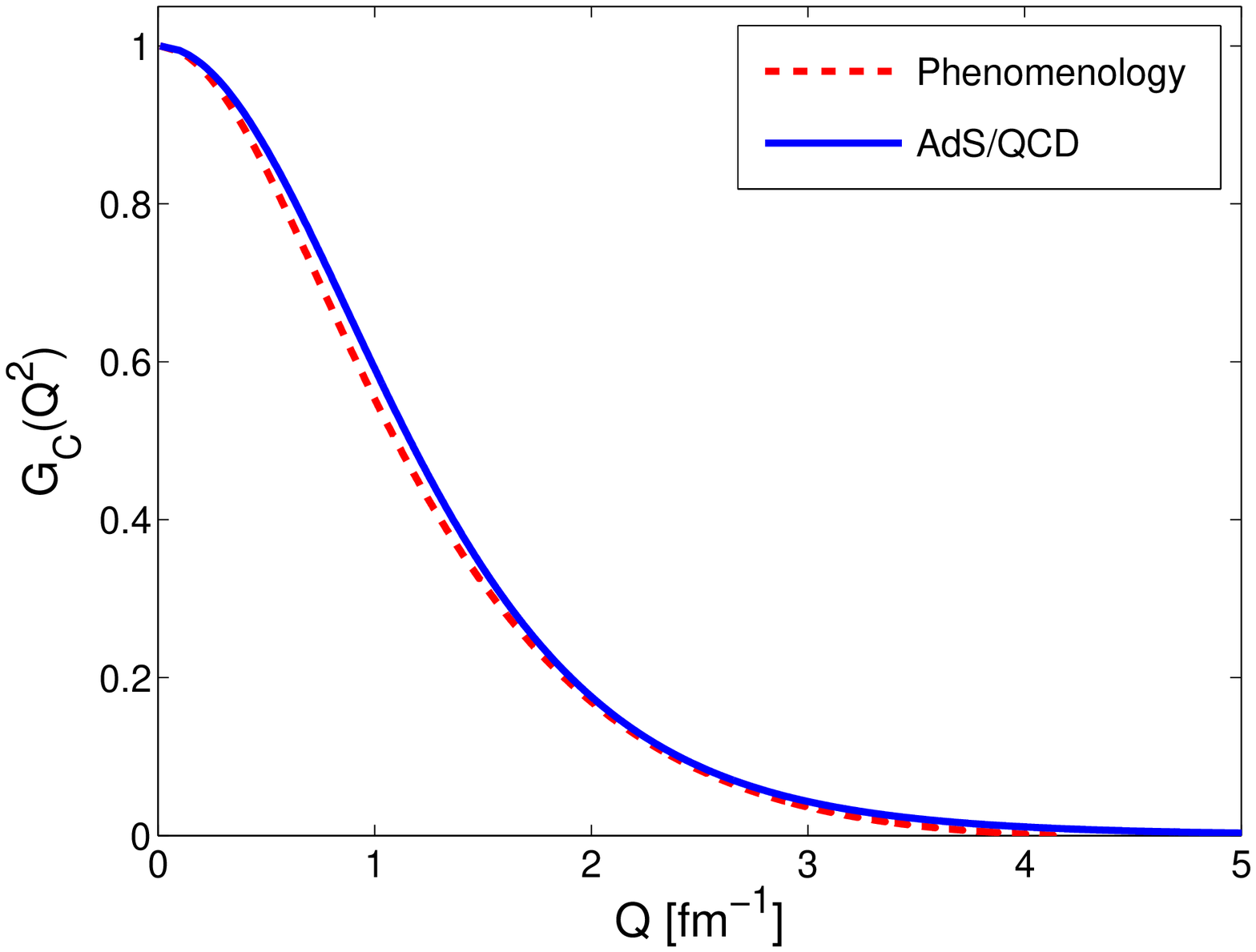}
{(b)}\includegraphics[width=5.4cm,clip]{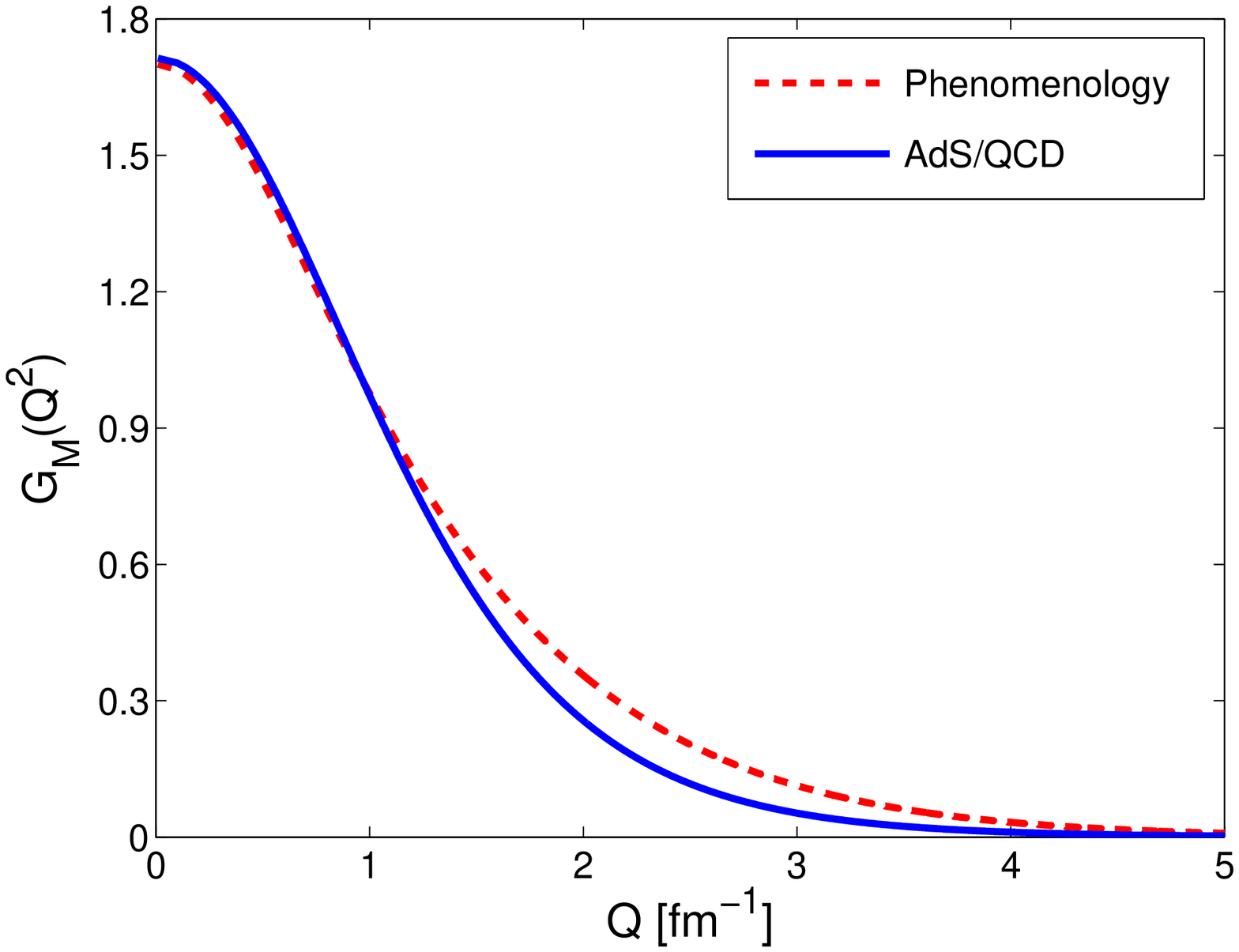}
{(c)}\includegraphics[width=5.4cm,clip]{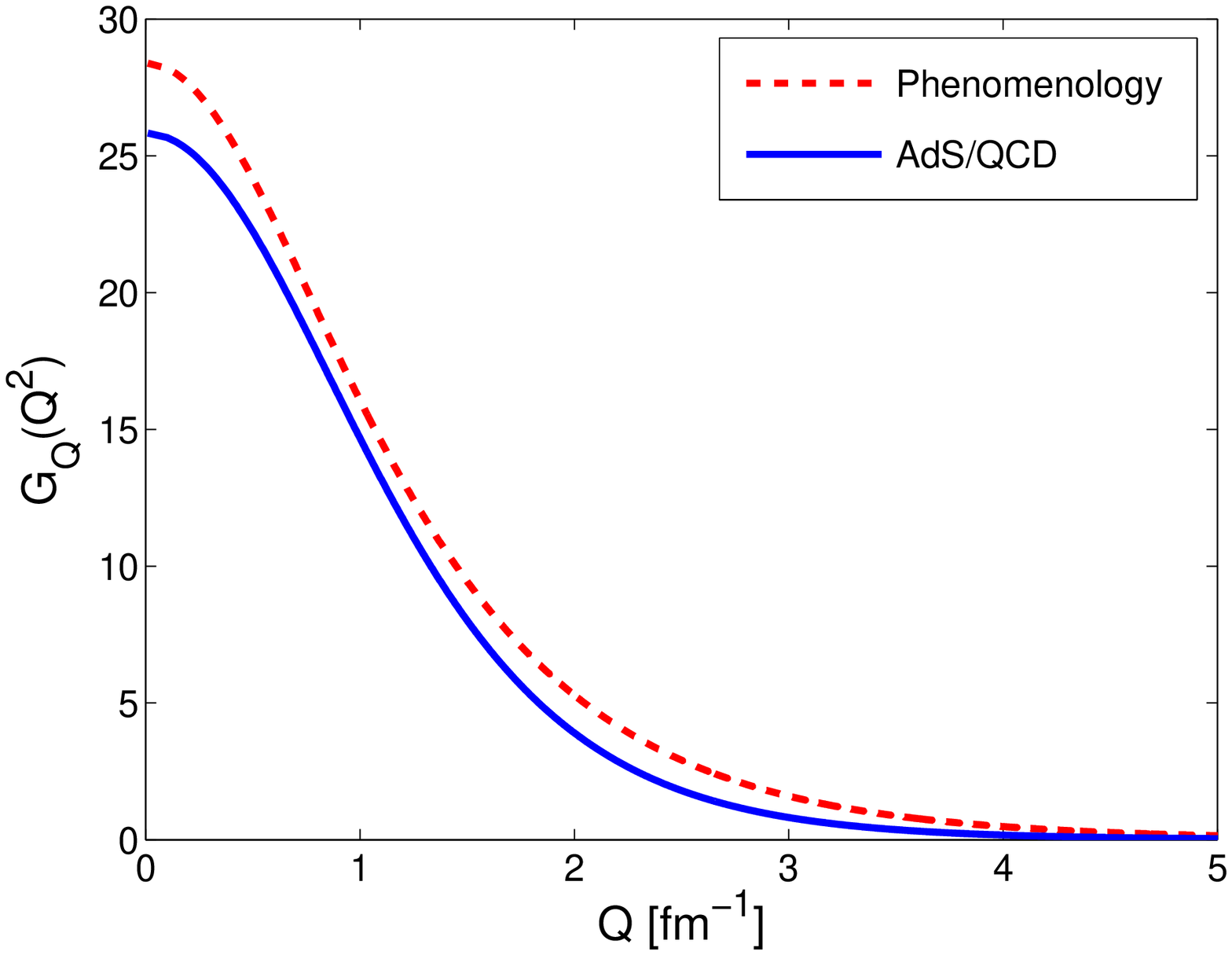}
\end{minipage}
\caption{\label{FFs}(Color online) Deuteron electromagnetic form factors (a) $G_{C}(Q^2)$, (b) $G_{M}(Q^2)$ and (c) $G_{Q}(Q^2)$ 
are plotted against $Q~fm^{-1}$. The red dashed line represents the phenomenological parameterization of the deuteron form factors data 
\cite{Abbott:2000ak} and the solid blue line represents the results of soft-wall AdS/QCD.} 
\end{figure*}

For the derivation of the deuteron electromagnetic
form factors in the framework of soft-wall AdS/QCD we follow the works of Gutsche $et~al.$ \cite{Gutsche1,Gutsche2,Lyubo}. 
For a hadronic bound state, the twist which is defined as dimension $-$ spin,  can also be written as $\tau=n_p$ where $n_p$ is the number of partons in a bound state \cite{deTeramond:2010ge,BT01,BT011,BT_new3}. The ground state of the deuteron which is a bound state of a proton and a neutron having six partons  is described by $\tau=6$ wavefunctions.
The effective action in terms of the AdS fields $d^M(x,z)$ and $V^M(x,z)$ which are dual to 
the Fock component contributing to the deuteron with twist $\tau = 6$ and the electromagnetic field, 
respectively, is given by \cite{Gutsche1}
\be\label{Eff_action} 
&&S = \int d^4xdz \, e^{-\varphi(z)} \, 
\Bigl[ - \frac{1}{4} F_{MN} F^{MN} 
- D^M d^\dagger_{N} D_M d^N  \nonumber\\
&&- i c_2 F^{MN} d^\dagger_{M} d_{N}
+ \frac{c_3}{4M_d^2} \, e^{2A(z)} \, \partial^M F^{NK}  
\Bigl( iD_K d^\dagger_{M} d_{N} \nonumber\\ 
&&- d^\dagger_{M} i D_K d_{N} + \mathrm{H.c.} 
\Bigr)  
+ 
d^\dagger_{M} \, \Big(\mu^2 +  U(z) \Big) \, d^M
\Bigr]  \,, 
\ee 
where 
$A(z) = \log(R/z)$, the stress tensor: $F^{MN}(x,z) = \partial^M V^N(x,z) - 
\partial^N V^M(x,z)$, $D^M$ is the covariant derivative, 
$\mu^2 R^2 = (\Delta - 1) (\Delta - 3)$ is the five-dimensional mass and 
$R$ is the AdS radius. The background dilaton field  $\varphi(z) = \kappa^2 z^2$. For a state
with $n_p$ partons and $L$ orbital angular momentum, the scaling dimension is defined as
$\Delta=n_p+L$ \cite{deTeramond:2010ge,BT01,BT011}. However, $n_p=\tau$ which corresponds to the fact that the dimension of the $d^M(x,z)$ field is $\Delta = \tau + L$. 
The first and second terms in Eq.\ref{Eff_action} correspond to kinematic parts of the electromagnetic and the deuteron fields respectively whereas the last part of the action denotes the effective mass term.
The rest of the terms in the effective action is due to the interaction which are responsible for generating the deuteron form factors.
The pure  AdS geometry reproduces 
the kinematical aspects of the light-front
Hamiltonian whereas the breaking of the maximal symmetry
of AdS by the background dilaton field allows the confinement dynamics of the theory in physical
space-time. In general, the presence of the quadratic
covariant derivatives with a dilaton in the effective action leads to a mixture of kinematical and dynamical effects. 
The $z$-dependent effective mass $\mu_{eff}(z)=\mu^2 +  U(z)$ is chosen to cancel the interference term and keep the mass 
term $\mu^2$ independent of $z$ \cite{BT_new3,deTeramond:2013it}.
 $U(z)$ in the effective mass gives  the confinement potential with 
$ 
U(z)  =  \frac{\varphi(z)}{R^2} \, U_0 \, ,
$
where the constant $U_0$ is determined by the value of the deuteron
mass $M_d$, and the parameters $c_2$ and $c_3$ are obtained 
by normalization of the deuteron electromagnetic form factors. Using the Kaluza-Klein
decomposition for the vector AdS field dual to the deuteron 
\be 
d^\mu(x,z) 
= \exp\Big[\frac{\varphi(z)-A(z)}{2}\Big] \, 
\sum\limits_n d^\mu_{n}(x) \Phi_{n}(z) \, , 
\ee
where $\Phi_{n}(z)$ are their bulk profiles and  $d^\mu_{n}(x)$ is the tower of
the Kaluza-Klein fields dual to the deuteron fields with radial quantum number $n$ 
and twist-dimension $\tau = 6$, one can derivative the equation of motion for 
the bulk profile $\Phi_{n}(z)$ as \cite{Gutsche1}
\be\label{EOM}
\Bigl[ - \frac{d^2}{dz^2} + \frac{4(L + 4)^2 - 1}{4z^2} 
+ \kappa^4 z^2 &+& \kappa^2 U_0 \Bigr]\Phi_{n}(z) \nonumber\\ 
&=& M_{d,n}^2 \Phi_{n}(z) \;.
\ee
This is a Schr\"odinger-type equation and the solutions of the Eq.(\ref{EOM}) are given by
\be
\Phi_{n}(z) &=& \sqrt{\frac{2n!}{(n+L+4)!}}  \, 
\kappa^{L+5} \, z^{L+9/2} \nonumber\\ 
&\times&\, e^{-\kappa^2 z^2/2} 
\, L_n^{L+4}(\kappa^2z^2)\,, \label{wf}\\
M_{d,n}^2 
&=& 4\kappa^2 \Bigl[ n + \frac{L + 5}{2} + \frac{U_0}{4}\Bigr]\,,  
\ee  
where $L_n^m(x)$ are the generalized Laguerre polynomials. For the ground state i.e. $n=0, \ L=0$, one gets 
$M_d = 2 \kappa \, \sqrt{\frac{5}{2} + \frac{U_0}{4}}$. 
Using the data for the deuteron mass $M_d = 1.875613$ GeV 
 the scale parameter $\kappa = 190$ MeV which is fixed by fitting the electromagnetic
form factors of the deuteron with experimental data, the constant $U_0$ is found to be $ 87.4494$ \cite{Gutsche1}.
In the soft-wall AdS/QCD framework, the deuteron electromagnetic form factors $G_{1,2,3}(Q^2)$ can be expressed 
in terms of the universal form factor $F(Q^2)$ such 
as \cite{Gutsche1,Gutsche2}
\be
G_1(Q^2) &=& F(Q^2)\, \label{G1},\\
G_i(Q^2) &=& c_iF(Q^2)\, , \quad\hspace*{.25cm} \{i=2,3\}\label{Gi}
\ee
where $F(Q^2)$ is the twist-6 hadronic form factor, 
which is given by the overlap of the square of the bulk profile dual to
the deuteron wave function and the confined electromagnetic current
\be\label{FQ21} 
F(Q^2) = \int\limits_0^\infty dz \, \Phi^2_0(z) \, 
V(Q,z) \,.
\ee  
$V(Q,z)$ is the vector bulk-to-boundary propagator dual to the $Q^2 (=-q^2)$-dependent electromagnetic
current and the integral representation of $V(Q,z)$ is given by \cite{Grigoryan:2007my,BT2},   
\be
\label{VInt}
V(Q,z) = 
\kappa^2 z^2 \int\limits_0^1 \frac{dx}{(1-x)^2} \, 
e^{-\kappa^2z^2 x/(1-x)} \, x^a \,. 
\ee
Using the deuteron wave function $\Phi_0(z)$ from Eq.(\ref{wf}) and $V(Q,z)$ from Eq.(\ref{VInt}), one finds $F(Q^2)$ as
\be\label{FQ2} 
F(Q^2) = \frac{\Gamma(6) \, \Gamma(a+1)}{\Gamma(a+6)}\, ,
\ee 
with $a = Q^2/(4 \kappa^2)$. Note that the formula follows the the general form of the hadronic form factor with twist $\tau$:
$ 
F_\tau(Q^2) = \frac{\Gamma(\tau) \, \Gamma(a+1)}{\Gamma(a+\tau)} 
$ 
derived in \cite{BT011}. The parameters $c_2$ and $c_3$ in Eq.(\ref{Gi}) are determined by the 
normalization of the deuteron form factors as defined in \cite{Gutsche1}: 
$c_2 = G_M(0) = 1.714$, and $ c_3 = G_M(0) + G_Q(0) - 1 = 26.544$.
We show the deuteron electromagnetic form factors $G_C(Q^2)$, $G_M(Q^2)$ and $G_Q(Q^2)$ obtained in the soft-wall the AdS/QCD
model in Fig.\ref{FFs}. The results are compared with the phenomenological parameterization of the experimental 
data of the deuteron form factors \cite{Abbott:2000ak}. One observes that $G_C$ evaluated in AdS/QCD
framework are in excellent agreement with the parameterization and $G_M$ and $G_Q$ in soft-wall AdS/QCD model are also  in more or less agreement with the parameterization. 
\begin{figure*}[htbp]
\begin{minipage}[c]{0.98\textwidth}
{(a)}\includegraphics[width=7.5cm,clip]{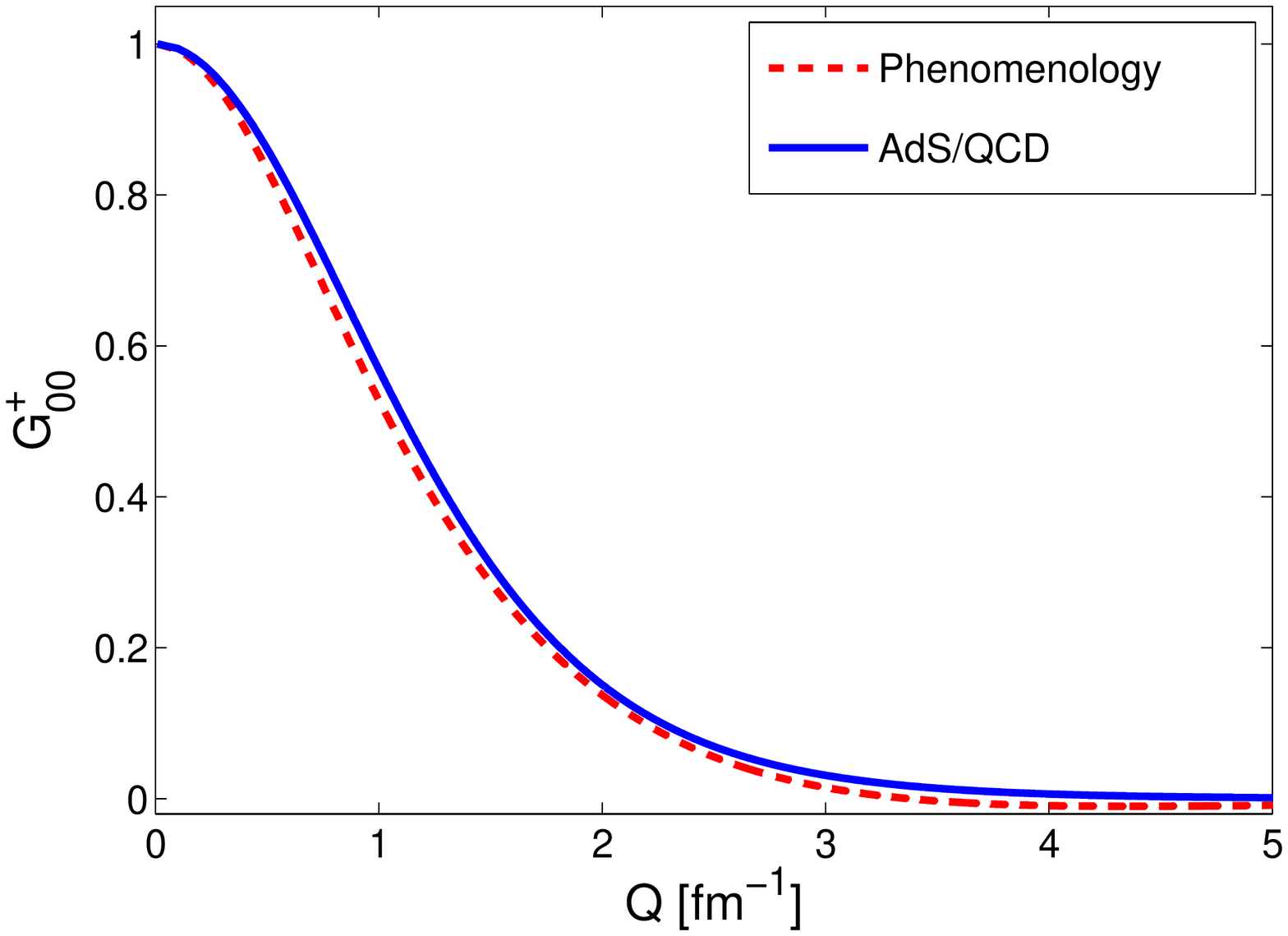}
{(b)}\includegraphics[width=7.5cm,clip]{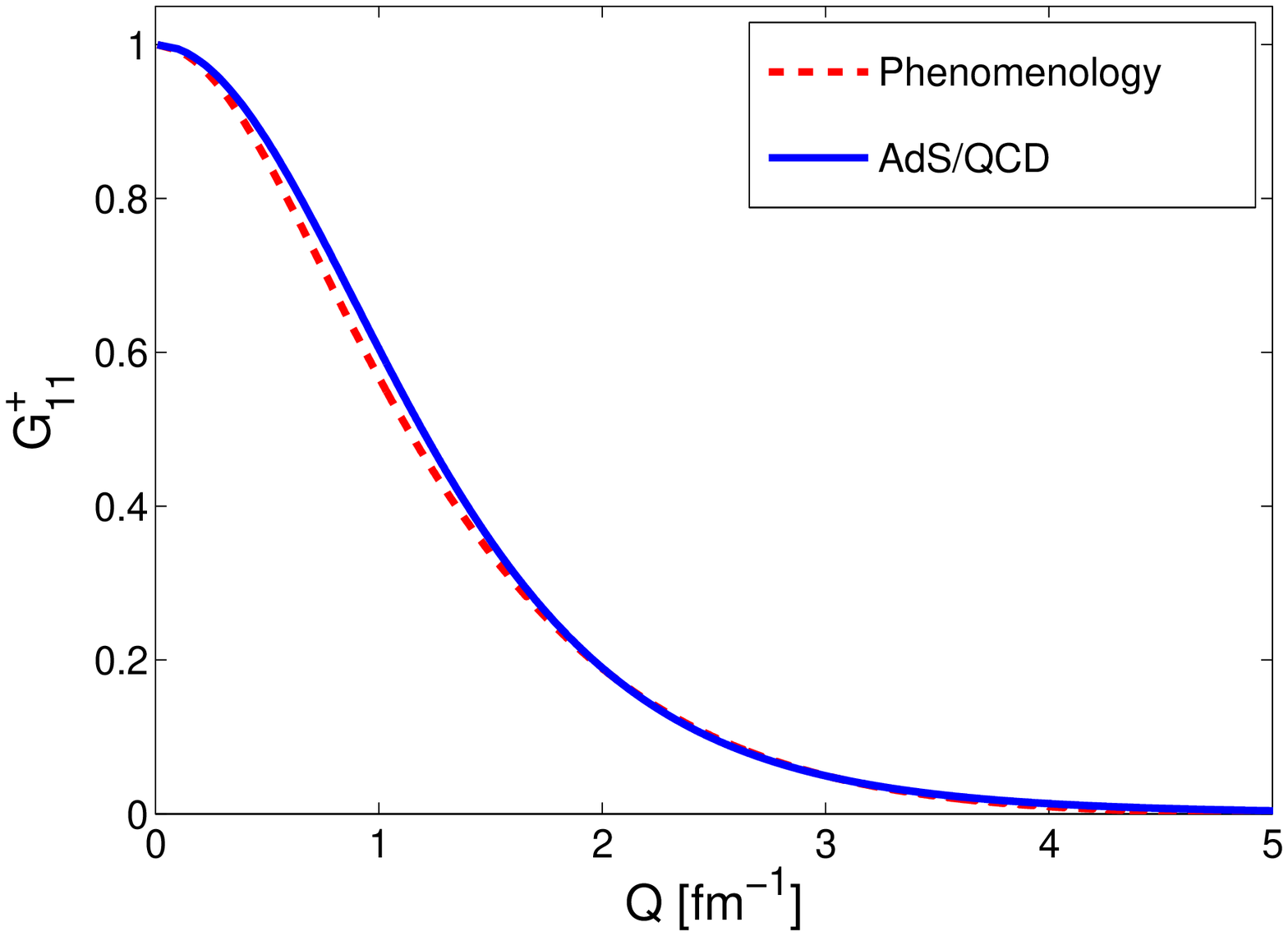}
\end{minipage}
\begin{minipage}[c]{0.98\textwidth}
{(c)}\includegraphics[width=7.5cm,clip]{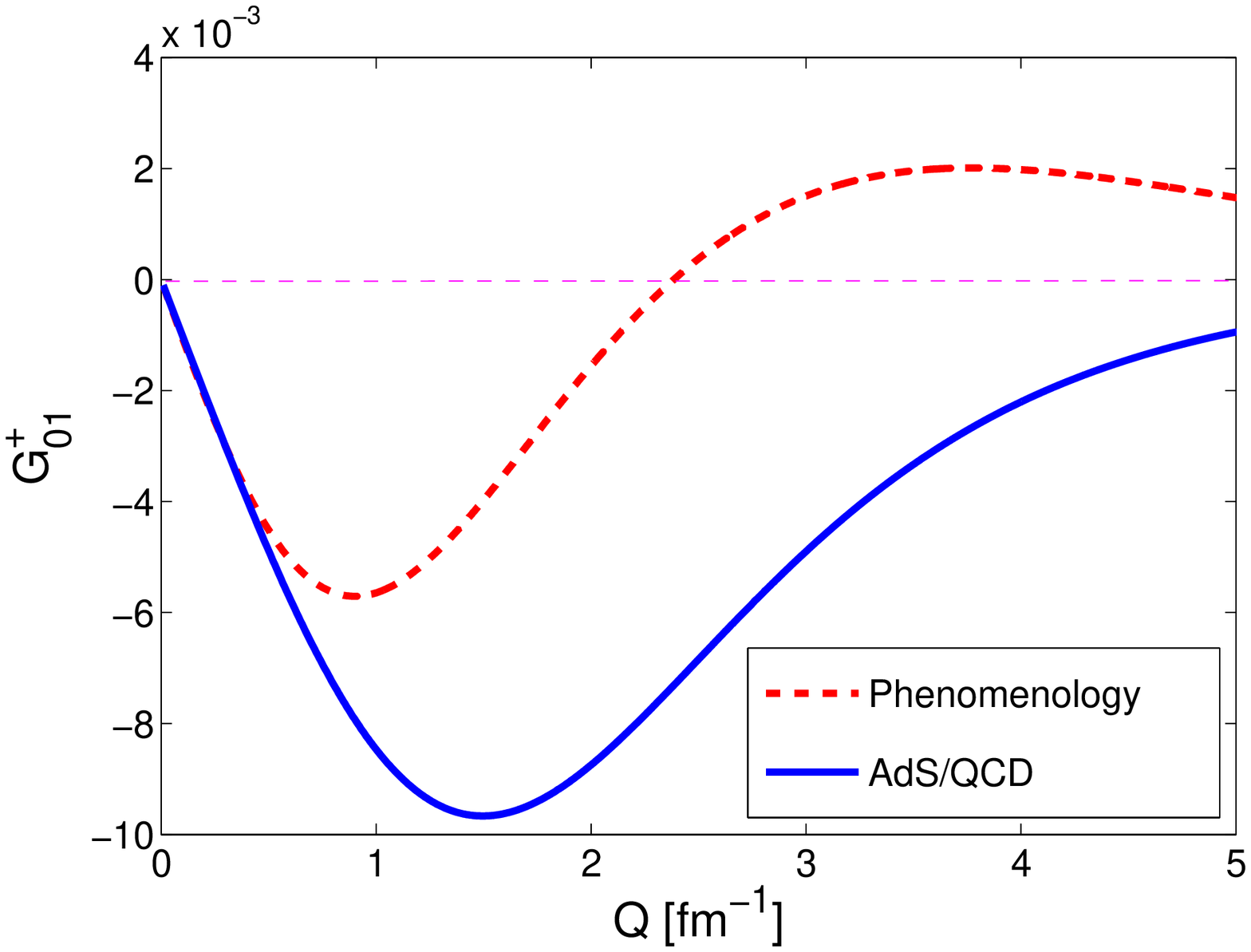}
{(d)}\includegraphics[width=7.5cm,clip]{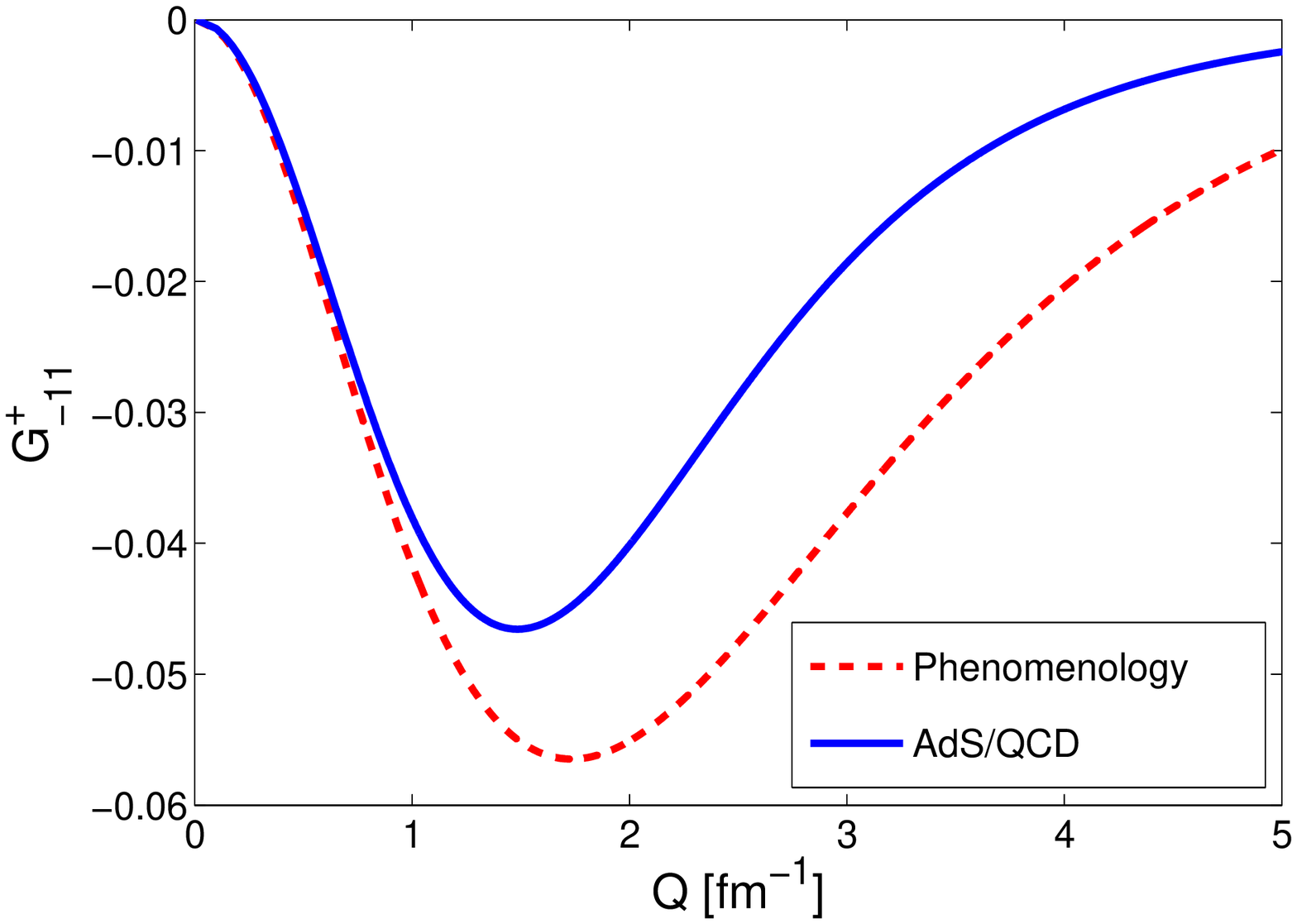}
\end{minipage}
\caption{\label{heli_con}(Color online) Deuteron helicity-conserving form factors (a) $G^+_{00}(Q^2)$, (b) $G^+_{11}(Q^2)$ and
helicity-flip form factors (c) with one unit of helicity-flip: $G^+_{01}(Q^2)$(a factor of $10^{-3}$ is taken out in the $y$-axis), (d) with two unit of helicity-flip: $G^+_{-11}(Q^2)$
are plotted against $Q$. The red dashed line represents the phenomenological parameterization of the deuteron form factors data 
\cite{Abbott:2000ak} and the solid blue line represents the results of soft-wall AdS/QCD.} 
\end{figure*}
\section{Transverse charge densities in the deuteron}\label{densities}
According to the standard interpretation \cite{miller07,vande,selyugin,weiss,CM3}, the charge density in the transverse
plane can be identified with the two-dimensional Fourier transform (FT) of the electromagnetic form factors in the 
light-cone frame with $q^+=q^0+q^3=0$. Thus, the transverse charge densities for the deuteron in helicity 
states of $\lambda = \pm 1$ or $\lambda = 0$ are defined as,
\begin{eqnarray}
\rho^d_{\lambda} (b) 
\!\!&\equiv& \!\!\int \frac{d^2 \mathbf{q}_\perp}{(2 \pi)^2} 
e^{- i \, \mathbf{q}_\perp \cdot \mathbf{b}_\perp} \, G^+_{\lambda \lambda}(Q^2)\nonumber\\
\!\!&=&\!\! \int_0^\infty \frac{d Q}{2 \pi} Q \, 
J_0(b \, Q) \, G^+_{\lambda \lambda}(Q^2),
\label{eq:dens2}
\end{eqnarray}
where the form factor $G^+_{\lambda \lambda^\prime}(Q^2)$ is again related to the matrix elements of
the electromagnetic current $J^+(0)$ operator between deuteron states as,
\begin{eqnarray}
&&\langle P^+, \frac{{\mathbf q}_\perp}{2}, \lambda^\prime | J^+(0) | 
P^+, - \frac{{\mathbf q}_\perp}{2}, \lambda  \rangle \nonumber\\
&=& (2 P^+) \, e^{i  (\lambda - \lambda^\prime) \phi_q} 
\, G^+_{\lambda^\prime \, \lambda} (Q^2),
\label{eq:dens1}
\end{eqnarray}
with $\lambda$~=~$\pm 1,0$ ($\lambda^\prime$~=$~\pm 1, 0$) which denotes the initial (final) deuteron light-front helicity, 
and $\mathbf{q}_\perp = Q ( \cos \phi_q \,\hat e_x + \sin \phi_q \,\hat e_y )$, and the impact parameter $\mathbf{b}_\perp = b \, 
(\cos \phi_b \, \hat e_x + \sin \phi_b \, \hat e_y)$  denotes the position in the transverse plane from the center of momentum ({\it c.m.}) of the 
deuteron. For $\lambda = \pm 1$ or $\lambda = 0$, there are two independent helicity conserving form factors $G^+_{1 \, 1}$ and 
$G^+_{0 \, 0}$ which can be expressed in terms of $G_{C, M, Q}$ as \cite{Carlson:2008zc}, 
\begin{eqnarray}
G^+_{1 \, 1} &=& \frac{1}{1 + \eta} 
\left \{G_C + \eta \, G_M + \frac{\eta}{3} \, G_Q \right \} , \label{EqG11}
\\
G^+_{0 \, 0} &=& \frac{1}{1 + \eta} 
\left \{(1 - \eta) \, G_C + 2 \eta \, G_M - 
\frac{2 \eta}{3} (1 + 2 \eta) G_Q \right \} , \nonumber\\ \label{EqG00}
\end{eqnarray}
with $\eta \equiv Q^2 / (4 M_d^2)$, and $M_d$ is the deuteron mass. Using the formulas of $G_C$, $G_M$
and $G_Q$ in terms of $F(Q^2)$ in the soft-wall AdS/QCD model, one can rewrite the expression of the helicity conserving form factors as
\begin{eqnarray}
G^+_{1 \, 1} &=& \left \{1 + \eta c_3\right \} F(Q^2)\, ,
\\
G^+_{0 \, 0} &=& 
\left \{ \frac{1-\eta-2\eta^2}{1+\eta}+2\eta c_2-2\eta^2c_3\right \}F(Q^2)\, .
\end{eqnarray}
\begin{figure*}[htbp]
\begin{minipage}[c]{0.98\textwidth}
{(a)}\includegraphics[width=7.5cm,clip]{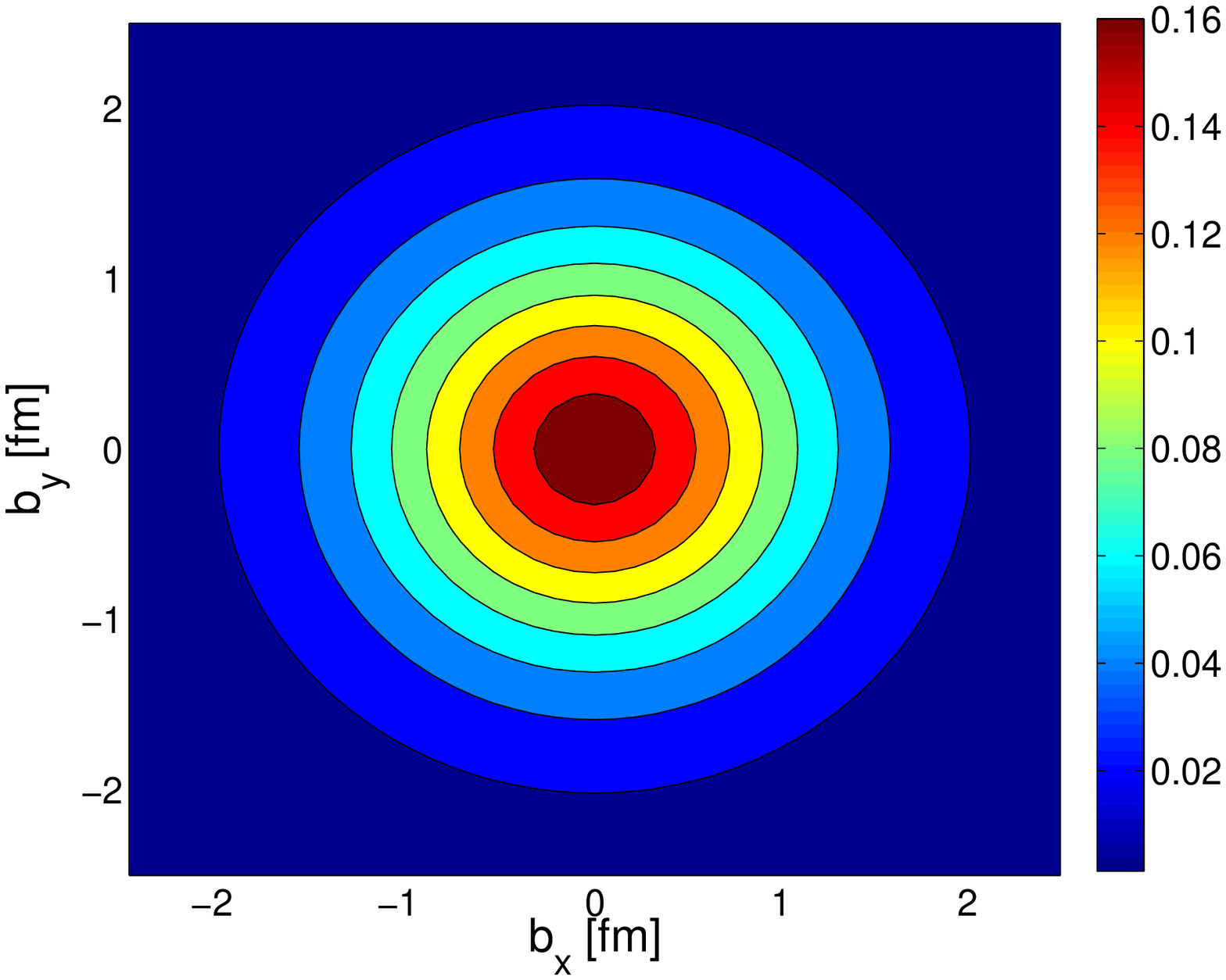}
{(b)}\includegraphics[width=7.5cm,clip]{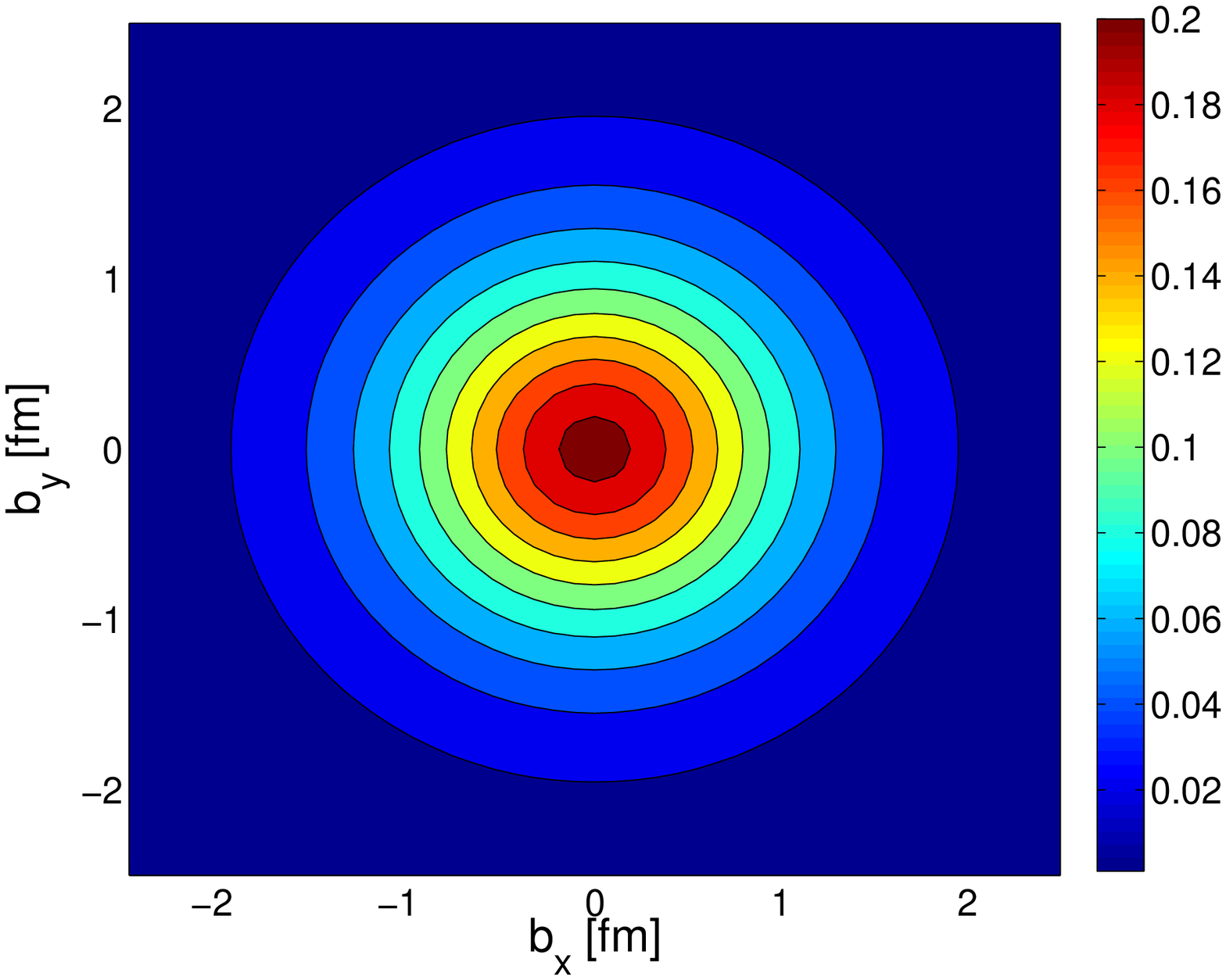}
\end{minipage}
\caption{\label{3D}(Color online) Top view of the three-dimensional quark transverse charge densities in the deuteron evaluated in AdS/QCD: (a) $\rho_0^d$ and (b) $\rho_1^d$ are for the longitudinally polarized deuterons.}
\end{figure*}
\begin{figure*}[htbp]
\begin{minipage}[c]{0.98\textwidth}
{(a)}\includegraphics[width=7.5cm,clip]{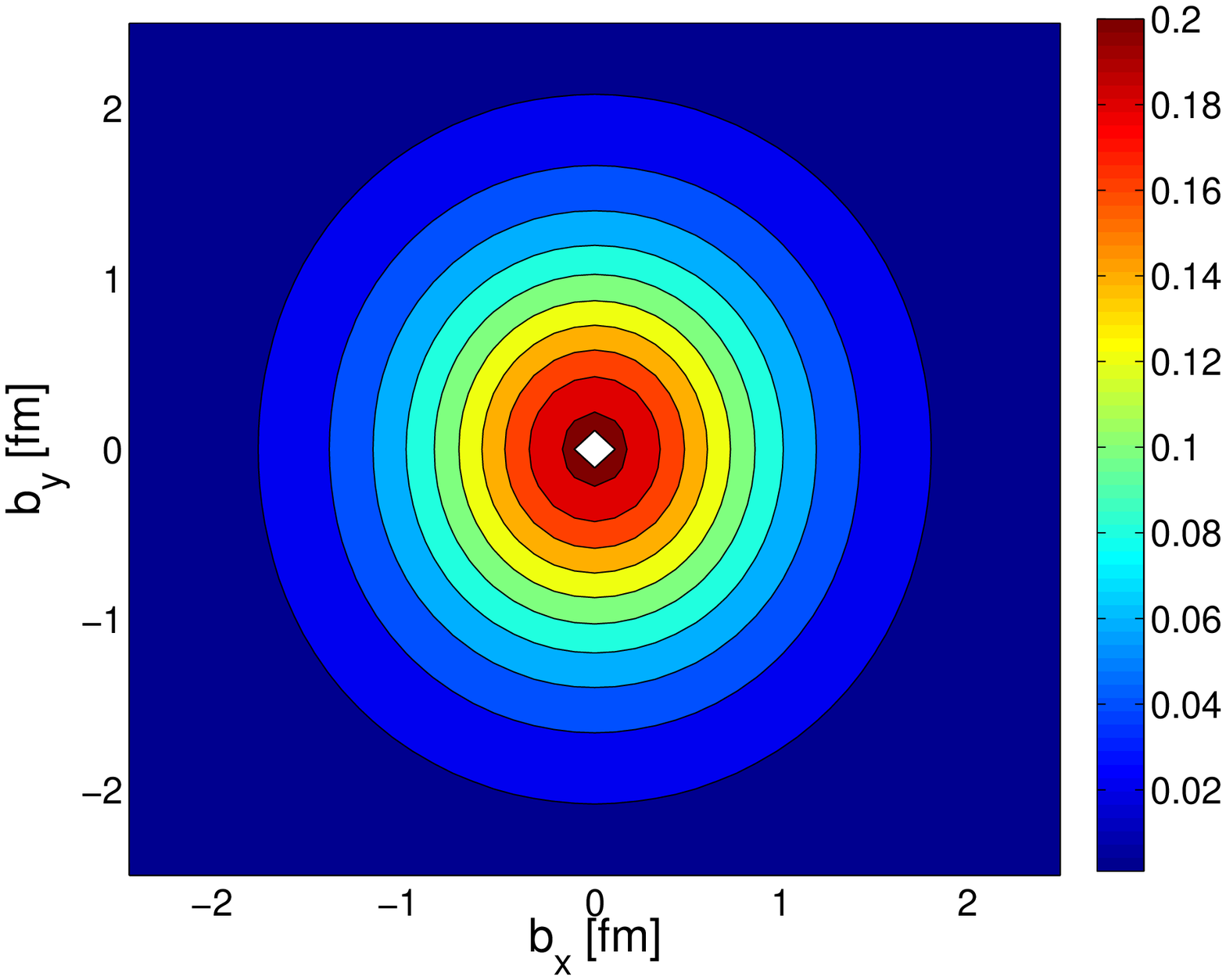}
{(b)}\includegraphics[width=7.5cm,clip]{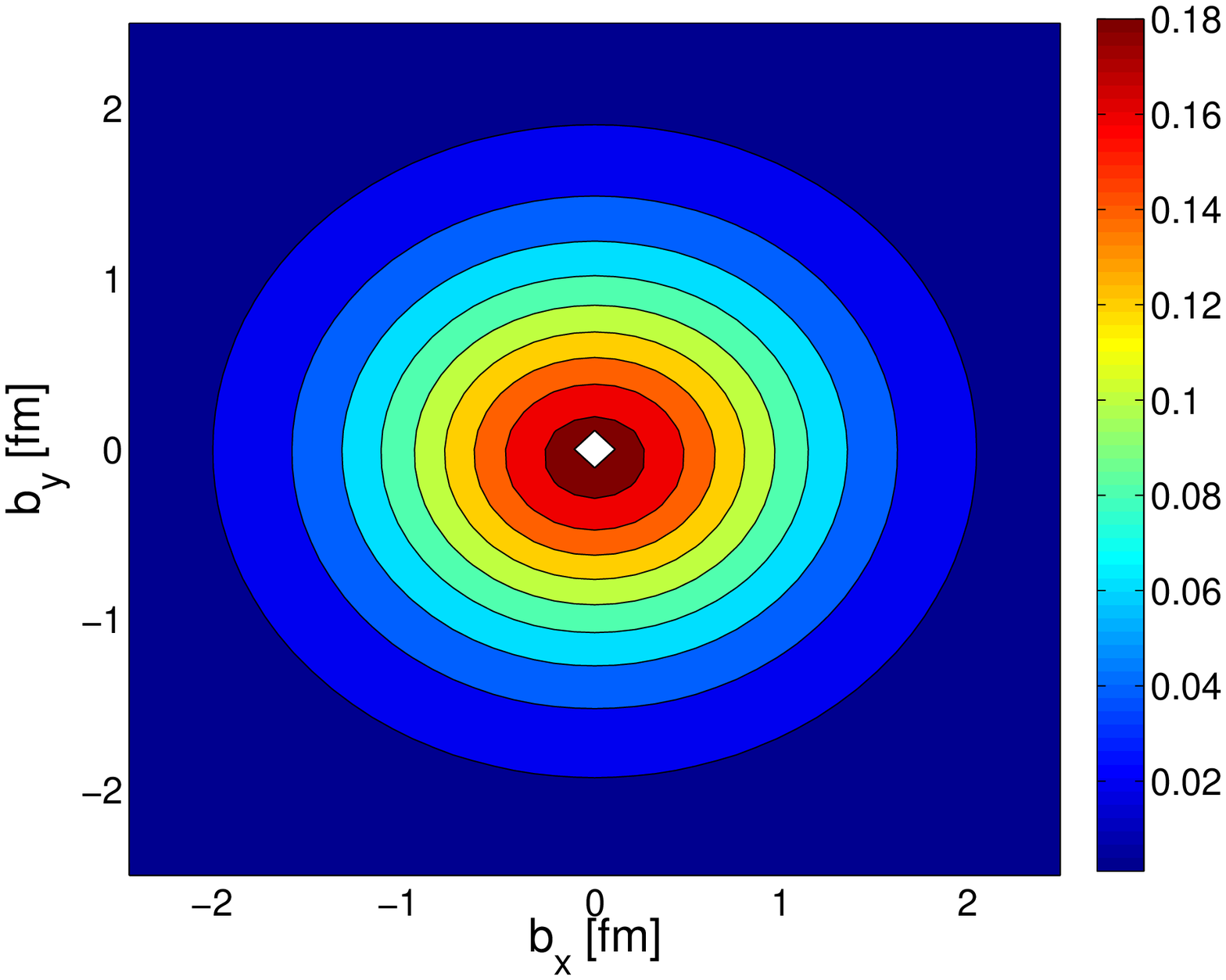}
\end{minipage}
\caption{\label{3DT}(Color online) Top view of the three-dimensional quark transverse charge densities in the transversely polarized deuteron evaluated in AdS/QCD: (a) $\rho_{T0}^d$ and (b) $\rho_{T1}^d$ are for the deuteron which is transversely polarized along the positive $x$-axis.}
\end{figure*}
In Fig.\ref{heli_con}-a and \ref{heli_con}-b, we show the deuteron helicity conserving form factors $G^+_{1 \, 1}$ and $G^+_{0 \, 0}$ respectively evaluated in the soft-wall AdS/QCD
model. In the same plots we compare the holographic results with the phenomenological parameterization of experimental 
data of deuteron form 
factors \cite{Abbott:2000ak}. One notices that $G^+_{1 \, 1}$ and $G^+_{0 \, 0}$ obtained in the AdS/QCD framework 
are in excellent agreement with the parameterization. 
The transverse charge densities for definite helicity i.e.
longitudinally polarized deuterons are shown in Fig.\ref{3D}. 
Both the charge densities for  $\lambda=0$ (Fig.\ref{3D}-a) and $\lambda=1$ (Fig.\ref{3D}-b) states are axially symmetric
and they have the peak at {\it c.m.}.

The transverse charge densities for longitudinally polarized deuteron lead to the monopole pattern only. To get information about dipole and the quadrupole moments of the deuteron states, one needs to consider the charge densities for the transversely polarized deuteron.
The transverse charge densities for the transversely polarized deuteron can be defined as,
\begin{eqnarray}
\rho^d_{T \, s_\perp} ({\mathbf b}) 
&\equiv& \int \frac{d^2 {\mathbf q}_\perp}{(2 \pi)^2} \,e^{- i \, {\mathbf q}_\perp \cdot {\mathbf b}} \,\nonumber\\ 
&\times& \frac{1}{2 P^+} 
\label{eq:dens4} 
\langle P^+, \frac{{\mathbf q}_\perp}{2}, s_\perp 
| J^+ | P^+, \frac{- {\mathbf q}_\perp}{2}, s_\perp  \rangle, 
\nonumber 
\end{eqnarray}
where $s_\perp$ is the deuteron  
spin projection along the transverse polarization direction, ${\mathbf S}_\perp = \cos \phi_S \,\hat e_x + \sin \phi_S \,\hat e_y$. 
For $s_\perp = +1$ and $s_\perp = 0$, the charge densities are given by \cite{Carlson:2008zc},
\begin{eqnarray}
\label{eq:dens5} 
\rho^d_{T \, 1} ({\mathbf b}) & = & \int_0^\infty \frac{d Q}{2 \pi} \, 
Q\, \left\{ 
J_0(b \, Q) \, \frac{1}{2} \left( G^+_{1 \, 1} + G^+_{0 \, 0} \right)
\right. \nonumber \\  
&&\hspace{0.5cm} 
+ \sin(\phi_b - \phi_S) \, J_1(b \, Q) \sqrt{2} \,  G^+_{0 \, 1}  \\ \nonumber
&&\left. \hspace{0.5cm}- \cos 2 (\phi_b - \phi_S) \, J_2(b \, Q) 
\frac{1}{2} \, G^+_{-1 \, +1}  
\right\} , \\ 
\rho^d_{T \, 0} ({\mathbf b}) & = & \int_0^\infty \frac{d Q}{2 \pi} \, 
Q\, \left\{ J_0(b \, Q) \, G^+_{1 \, 1} \right. \nonumber \\  
&&\left. \hspace{0.5cm}+ \cos 2 (\phi_b - \phi_S) \, J_2(b \, Q) 
\, G^+_{-1 \, +1} \right\} . 
\label{eq:dens6}  
\end{eqnarray}
It can be noticed that $\rho^d_{T \, 1}$ is a linear combination of helicity conserving charge densities together with two other components.
The second term involves  one unit of light-front helicity flip ($0 \to 1$) deuteron form factor which gives a dipole field pattern in the 
charge density. The last term in $\rho^d_{T \, 1}$, involves the form factor with two unit of  helicity flip ($-1 \to +1$) 
and corresponds to a quadrupole field pattern in the charge density. But $\rho^d_{T \, 0}$ does not contain the dipole field pattern.
One writes the deuteron form factor with one unit of helicity flip in terms of $G_{C, M, Q}$ as \cite{Carlson:2008zc}, 
\begin{eqnarray}
G^+_{0 \, 1} &=& - \frac{\sqrt{2 \eta}}{1 + \eta} 
\left \{G_C - \frac{1}{2} (1 - \eta) \, G_M + \frac{\eta}{3} \,G_Q \right \} , \label{EqG01}
\end{eqnarray}
whereas the deuteron form factor with two units of helicity flip, 
which governs the quadrupole field patterns is given by, 
\begin{eqnarray}
G^+_{-1 \, +1} &=& \frac{\eta}{1 + \eta} 
\left \{G_C - G_M - (1 + \frac{2 \eta}{3}) \,G_Q \right \} . \label{EqG1m1}
\end{eqnarray}
In the soft-wall AdS/QCD model the deuteron helicity form factors with one and two units of helicity flip,
can be expressed in terms of $F(Q^2)$ as,
\begin{eqnarray}
G^+_{0 \, 1} &=& -\sqrt{2 \eta}\left \{1 - \frac{c_2}{2} + \eta c_3\right \} F(Q^2)\, ,
\\
G^+_{-1 \, +1} &=& 
 - \eta c_3 F(Q^2)\, .
\end{eqnarray}
We show the deuteron helicity flip form factors, $G^+_{0 \, 1}$ and $G^+_{-1 \, +1}$
calculated in AdS/QCD framework in Fig.\ref{heli_con}-c and Fig.\ref{heli_con}-d, respectively. It can be noticed that though the deuteron
helicity conserving form factors in the AdS/ QCD agree well with the phenomenological parameterization, the helicity flip form factors
deviate at higher values of $Q$.  
Since $\eta$ is very small within the range of $Q^2$ in the plots, for the spin non-flip form factors
(Eqs.\ref{EqG11} and \ref{EqG00}) the only dominating contribution is coming from $G_C$ which is in very good agreement 
with phenomenology (see Fig.\ref{FFs}-a), $G_M$ and $G_Q$ are suppressed by  $\eta$.  Thus, the spin conserving form factors 
 agree well with the phenomenological 
parameterization. But for spin-flip form factor $G^+_{01}$ (Eq.\ref{EqG01}), the major contributions are coming 
from $G_C$, and $G_M$ ($G_Q$ is suppressed by $\eta$) with a relative negative sign which causes the small magnitude of $G^+_{01}$ 
(of the order of $10^{-3}$) but the errors add up as can be seen from Fig.\ref{FFs}.
A similar reason is also applicable for the mismatch in $G^+_{-1+1}$.

\begin{figure*}[htbp]
\begin{minipage}[c]{0.98\textwidth}
{(a)}\includegraphics[width=7.5cm,clip]{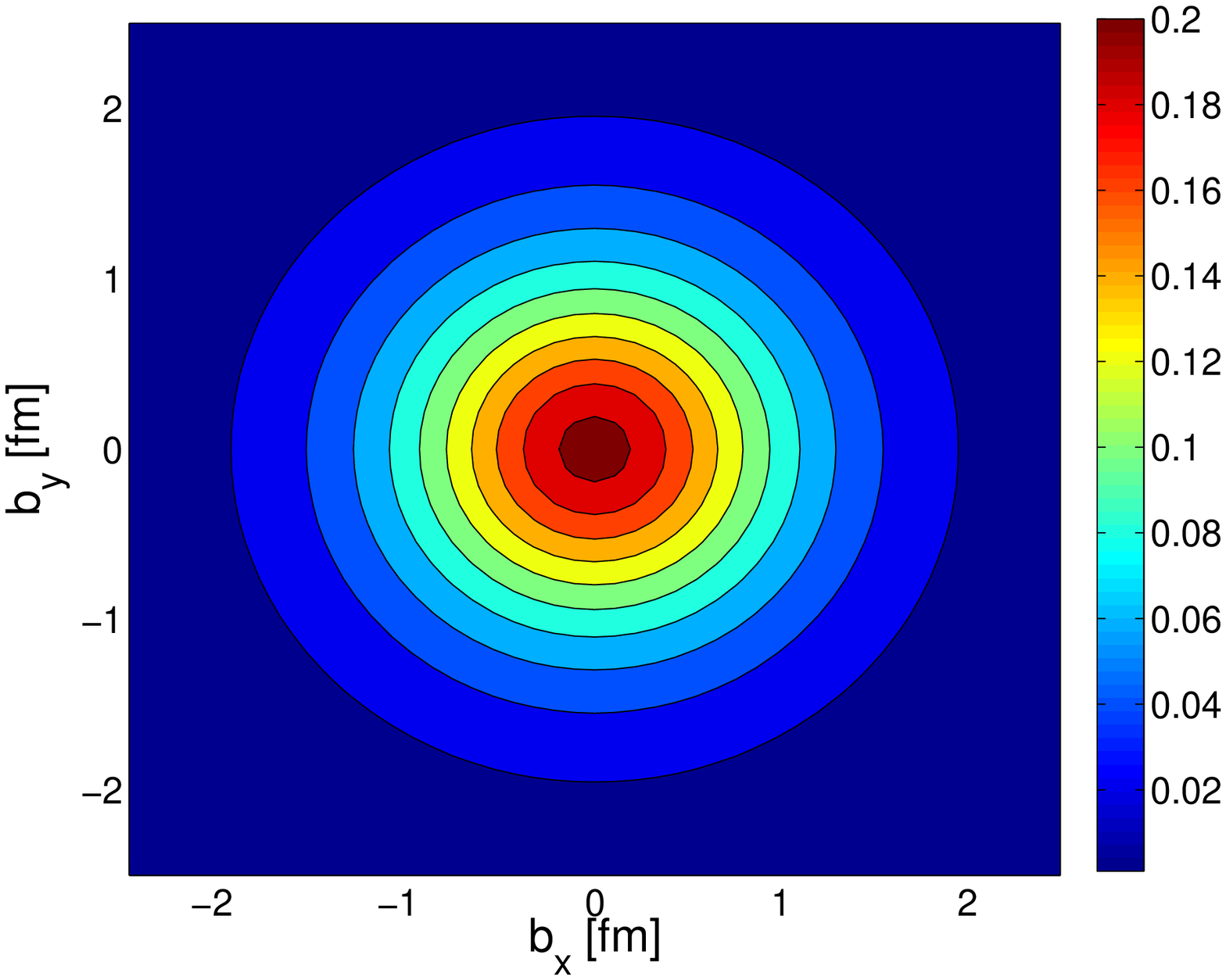}
{(b)}\includegraphics[width=7.5cm,clip]{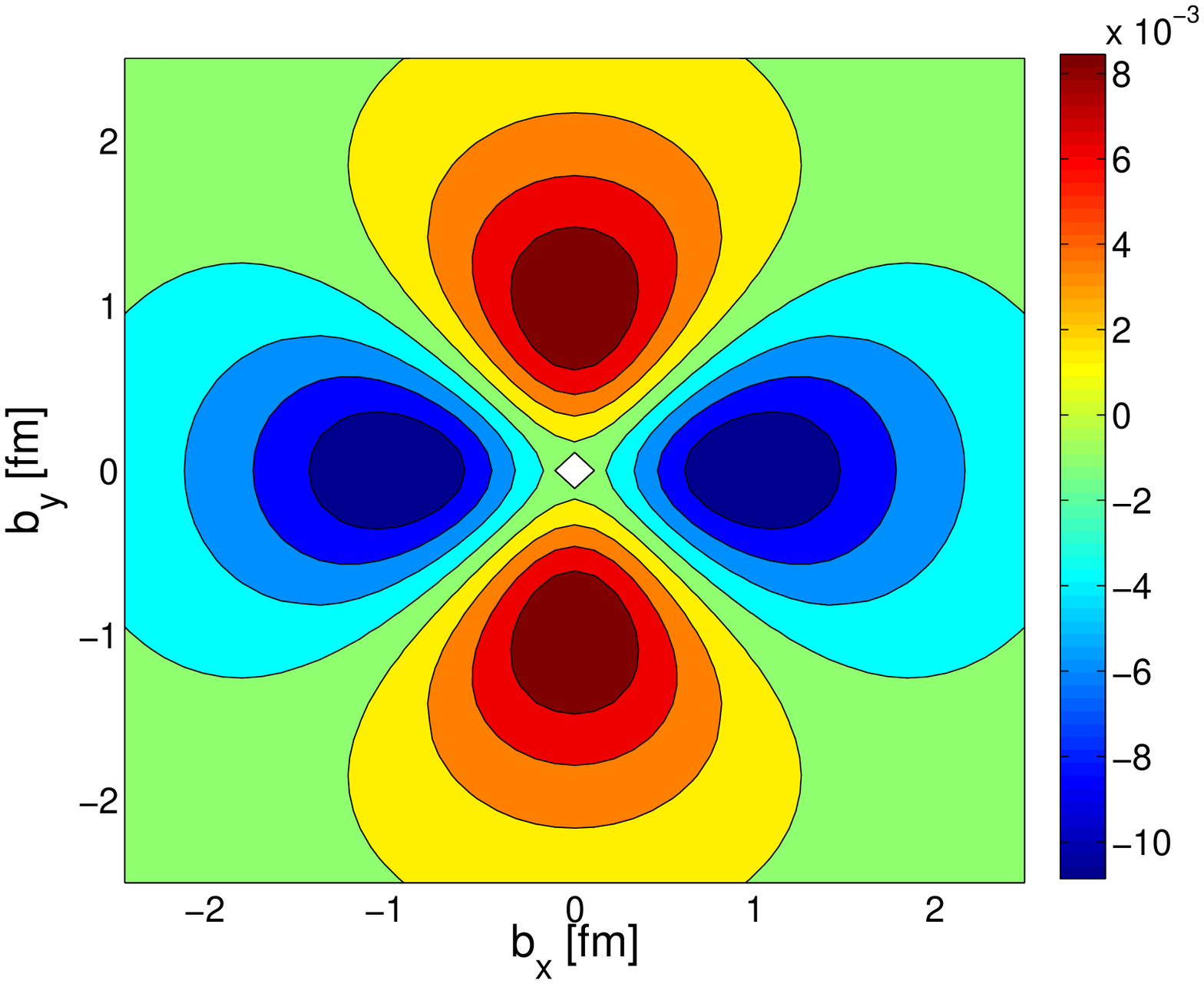}
\end{minipage}
\caption{\label{r0}(Color online) (a) Monopole and (b) quadrupole contributions to the charge density: $\rho_{T0}^d$.}
\end{figure*}
\begin{figure*}[htbp]
\begin{minipage}[c]{0.98\textwidth}
{(a)}\includegraphics[width=5.4cm,clip]{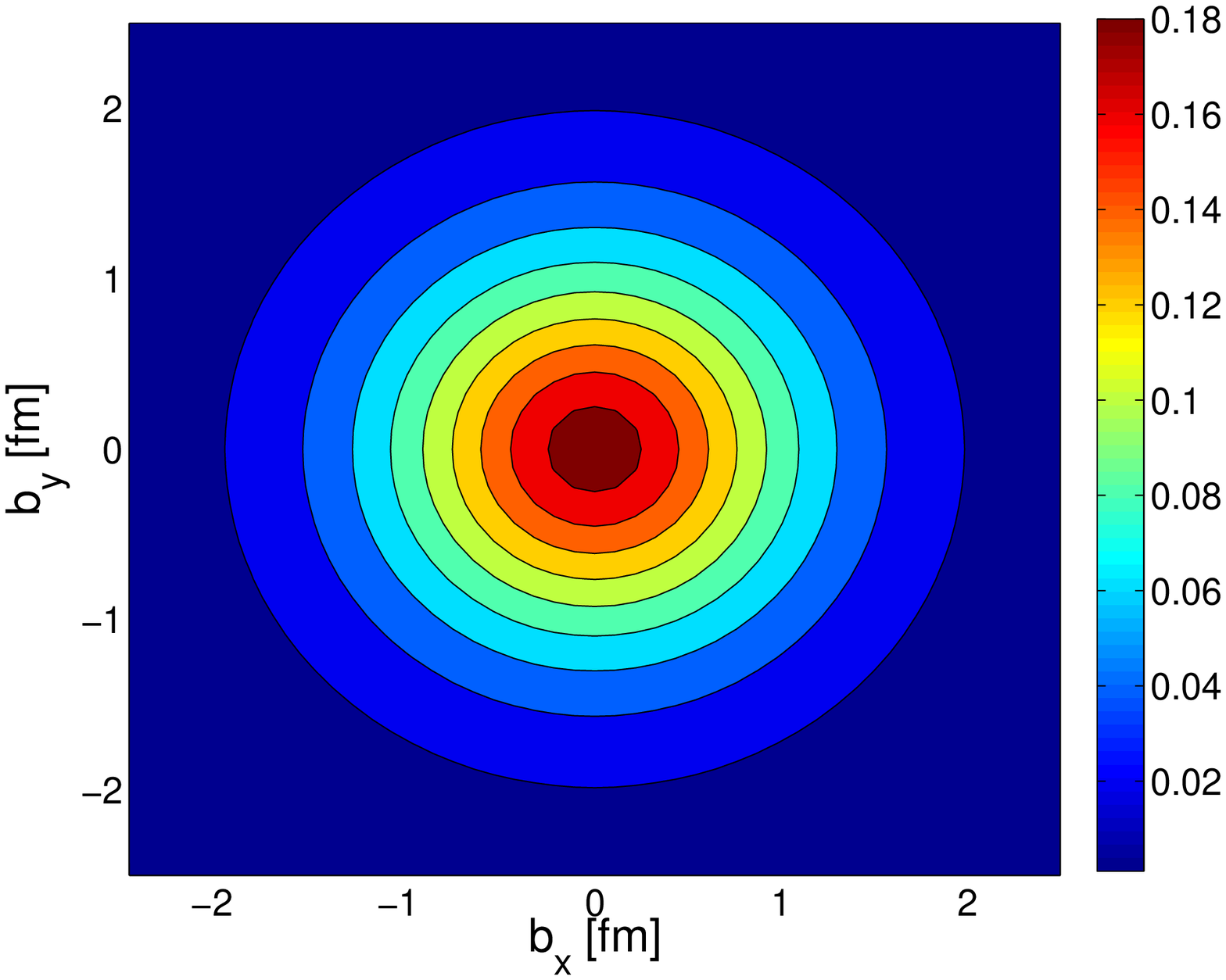}
{(b)}\includegraphics[width=5.4cm,clip]{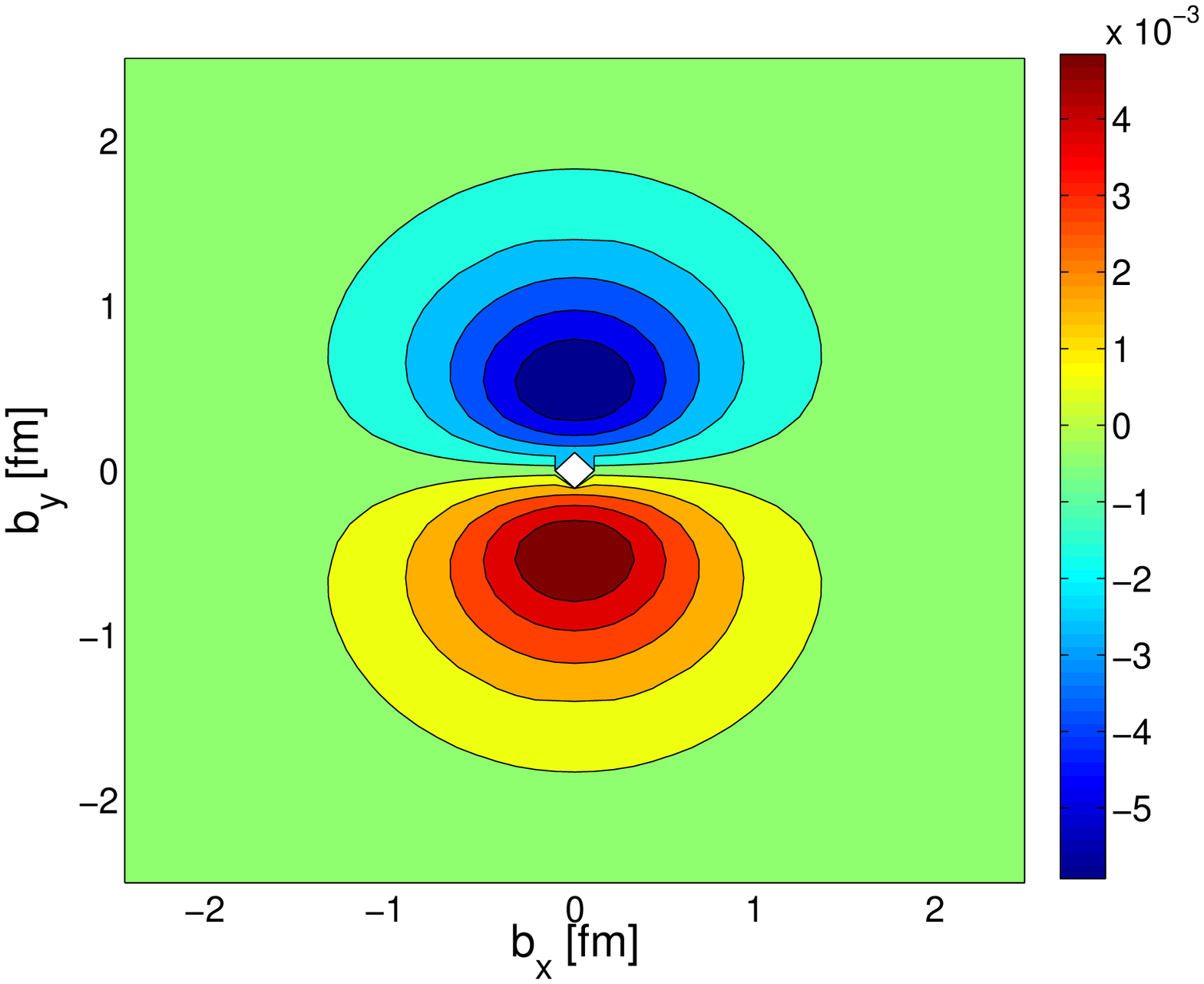}
{(c)}\includegraphics[width=5.4cm,clip]{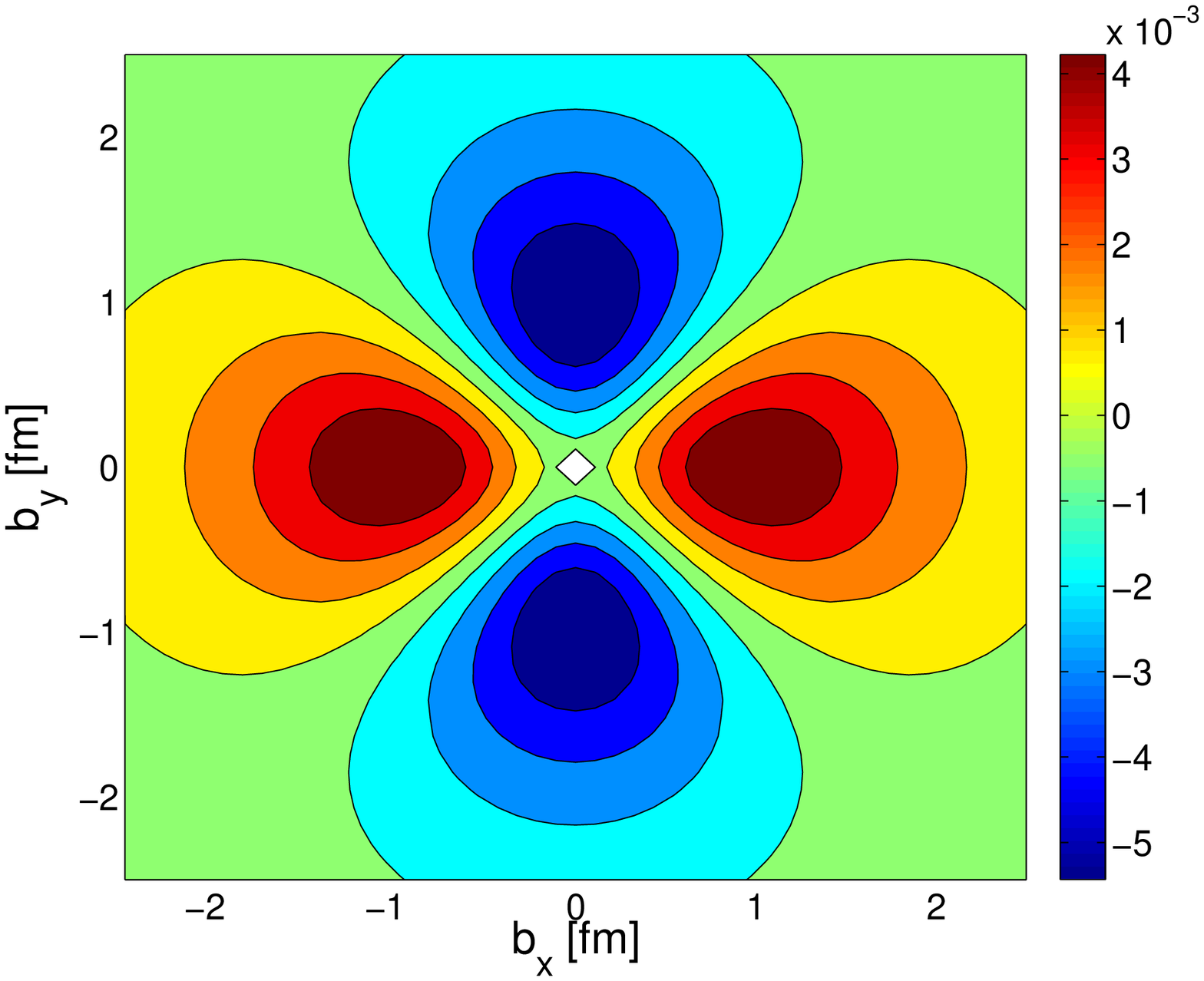}
\end{minipage}
\caption{\label{rT}(Color online) (a) Monopole, (b) dipole and (c) quadrupole contributions to the charge density: $\rho_{T1}^d$.}
\end{figure*}
\begin{figure*}[htbp]
\begin{minipage}[c]{0.98\textwidth}
{(a)}\includegraphics[width=7.5cm,clip]{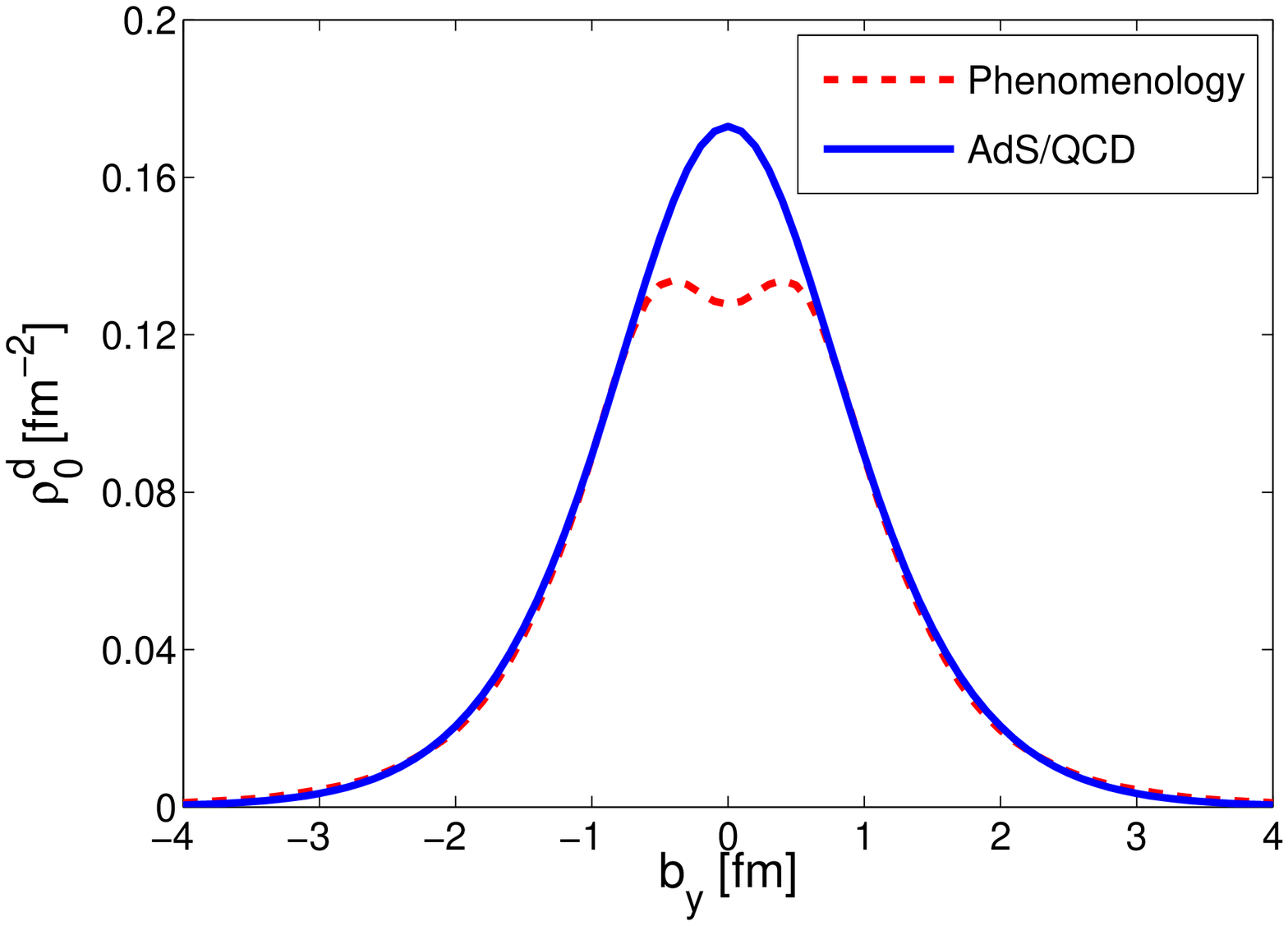}
{(b)}\includegraphics[width=7.5cm,clip]{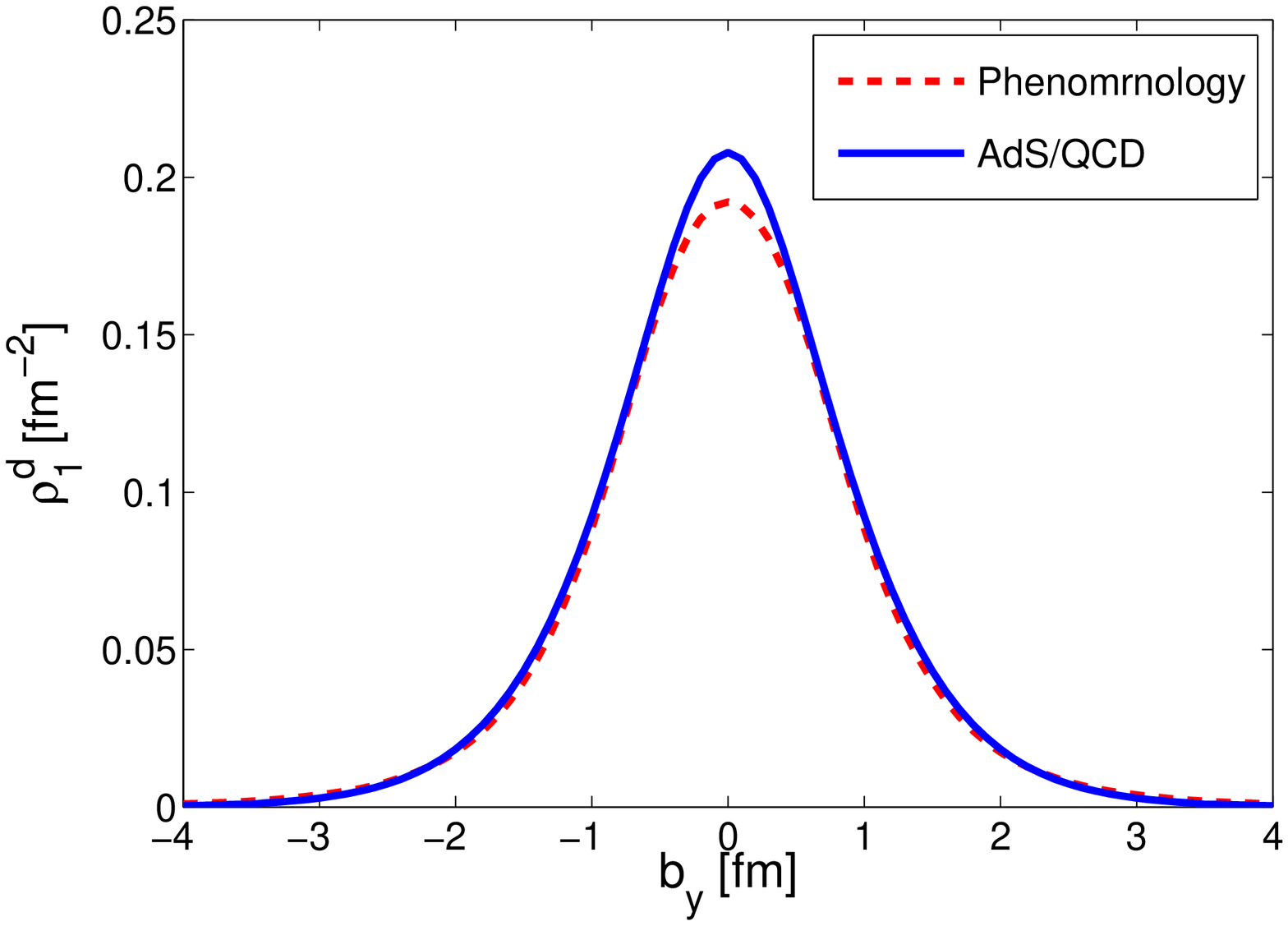}
\end{minipage}
\begin{minipage}[c]{0.98\textwidth}
{(c)}\includegraphics[width=7.5cm,clip]{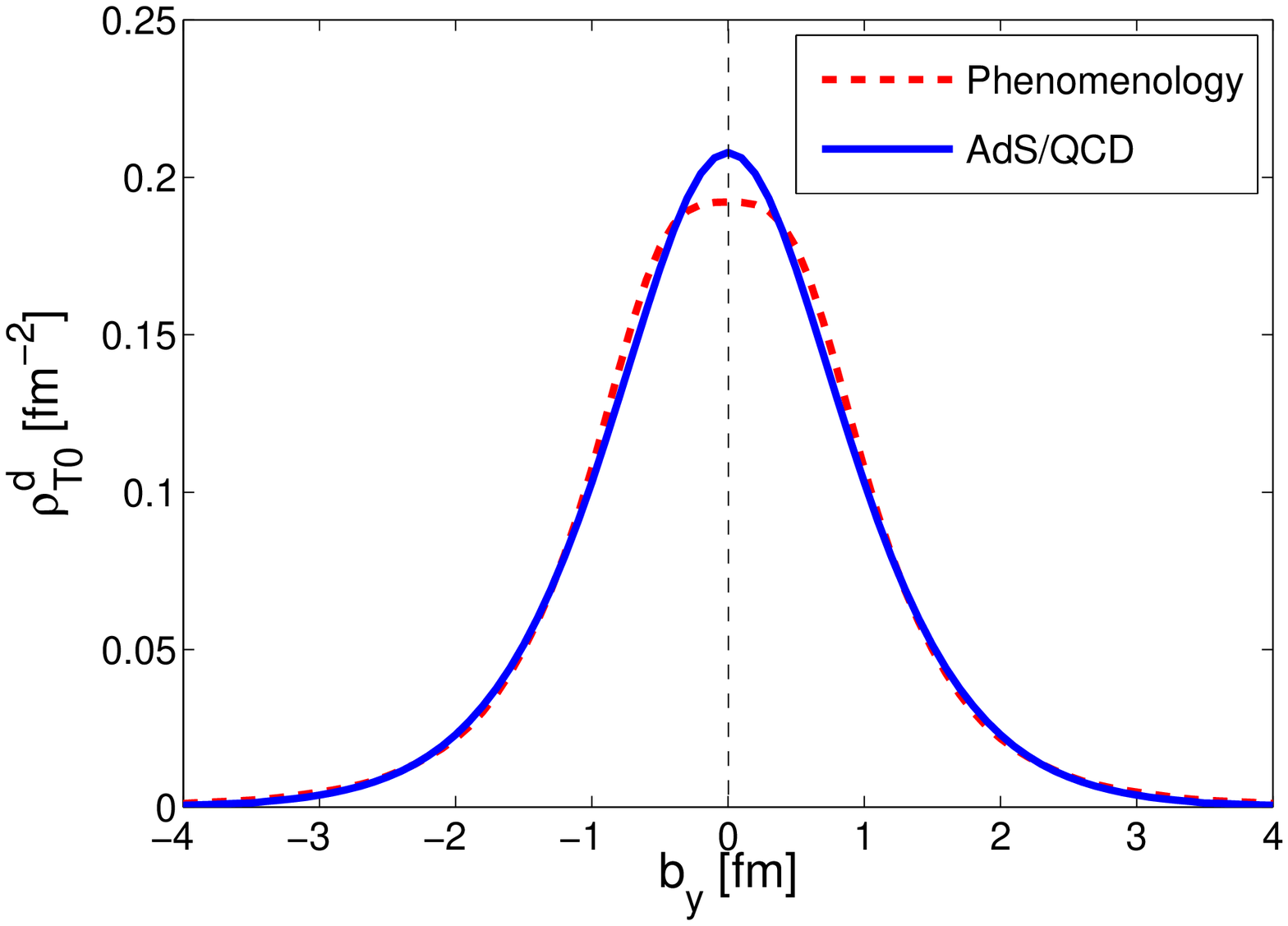}
{(d)}\includegraphics[width=7.5cm,clip]{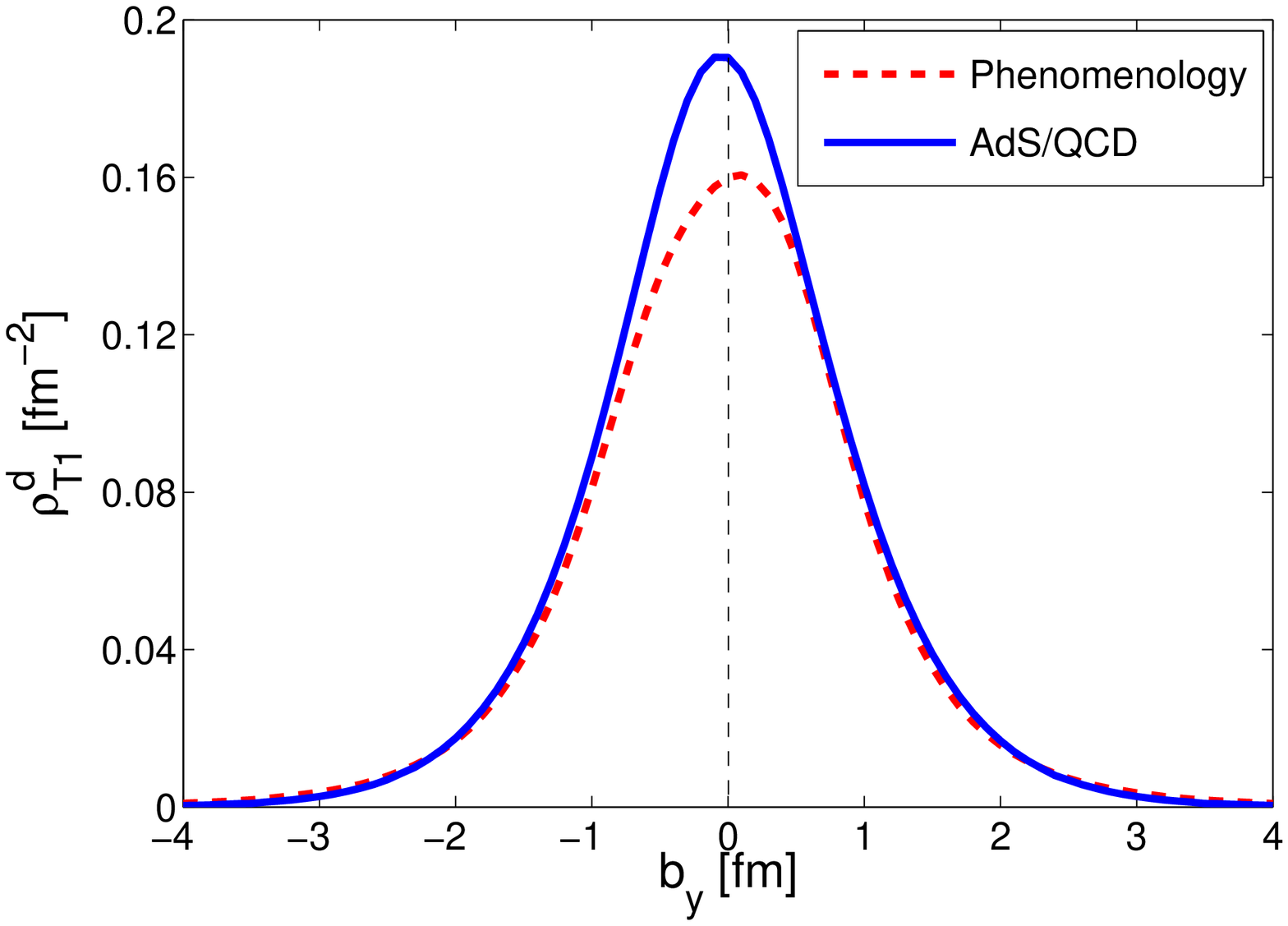}
\end{minipage}
\caption{\label{2D}(Color online) Comparison of transverse charge densities in the deuteron evaluated in AdS/QCD and the phenomenological results: (a) $\rho_0^d$ and (b) $\rho_1^d$ are for the longitudinally polarized deuterons (c) $\rho_{T0}^d$ and (d) $\rho_{T1}^d$ are for the deuteron which is transversely polarized along the positive $x$-axis. The solid blue and red dashed lines represent the results of AdS/QCD and the phenomenological results \cite{Carlson:2008zc} respectively.}
\end{figure*}
The AdS/QCD results for the transverse charge density with transverse deuteron
polarization are shown in Fig.\ref{3DT}. Fig.\ref{3DT}-a shows the charge distribution for the state with spin
projection-0 in the $x$-direction whereas Fig.\ref{3DT}-b exhibits charge density for a state with spin projection-1 in 
the $x$-direction. One finds that the effects of the quadrupole term in $\rho^d_{T \, 0}$  stretches the charge along 
the $y$-axis but does not cause any shift of the charge center whereas the quadrupole term in $\rho^d_{T \, 1}$ stretches the charge along 
the $x$-axis and the dipole term causes a small overall shift of the charge distribution 
along the $y$-axis. We show the individual monopole and quadrupole contributions to the charge density, $\rho^d_{T \, 0}$ in Fig.\ref{r0}
and the monopole, dipole, and quadrupole contributions to $\rho^d_{T \, 1}$ in Fig.\ref{rT}.
The dipolar and quadrupole patterns come in the densities due to the small anomalous magnetic moment coming from the second term of
the Eq.(\ref{eq:dens5}) which produces an induced electric dipole moment in $y$-direction and the anomalous electric quadrupole moment 
coming from the third term of the Eq.(\ref{eq:dens5}). 
It can be noted that the signs of
quadrupole contributions to the charge densities are opposite, thus, the stretching of $\rho^d_{T \, 0}$ is along the $y$-axis
but for $\rho^d_{T \, 1}$, it is along the $x$-axis. The dipole or quadrupole contributions to the charge densities are
comparatively very weak with respect to the monopole contributions. A comparison of the deuteron charge densities evaluated in soft-wall the 
AdS/QCD model and the phenomenological results \cite{Carlson:2008zc} obtained using the parameterization of the experimental data for 
the deuteron electromagnetic form factors \cite{Abbott:2000ak} are shown in Fig.\ref{2D}. We find that 
the AdS/QCD predictions for $\rho^d_{1}$ and $\rho^d_{T \, 0}$ are in good agreement with the phenomenological results 
except at $b=0$.  The dip appearing at $b=0$ in Fig.\ref{2D}(a) might be an artifact of the parameterization of the phenomenological data.
Again, the holographic prediction shows the overall small shift of $\rho^d_{T \, 1}$ from the {\it c.m.} in opposite 
direction to the phenomenological result. The shifting in $\rho^d_{T \, 1}$ occurs due to the dipole contribution which is coming from the one unit of light-front
helicity flip ($0 \rightarrow 1$) deuteron form factor, $G_{01}^+(Q^2)$. It can be noticed that the phenomenological parameterization for $G_{01}^+(Q^2)$ shows a positive value after $Q\sim 2.3~fm^{-1}$, however, it is always negative in the holographic QCD model (Fig.\ref{heli_con}-c). This produces the opposite shift in $\rho^d_{T \, 1}$ in the AdS/QCD model from the phenomenological result. 
\begin{figure*}[htbp]
\begin{minipage}[c]{0.98\textwidth}
{(a)}\includegraphics[width=7.5cm,clip]{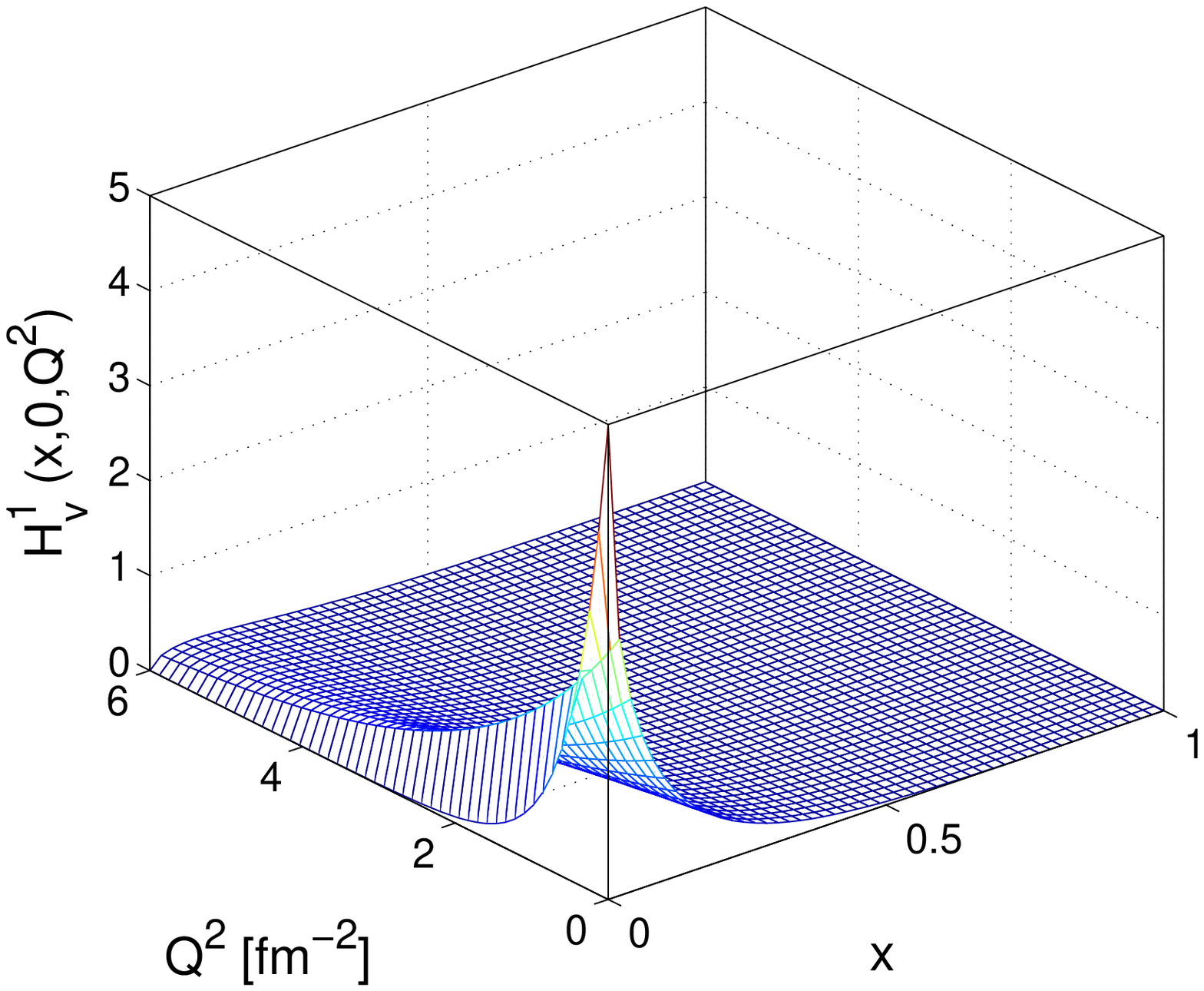}
{(b)}\includegraphics[width=7.5cm,clip]{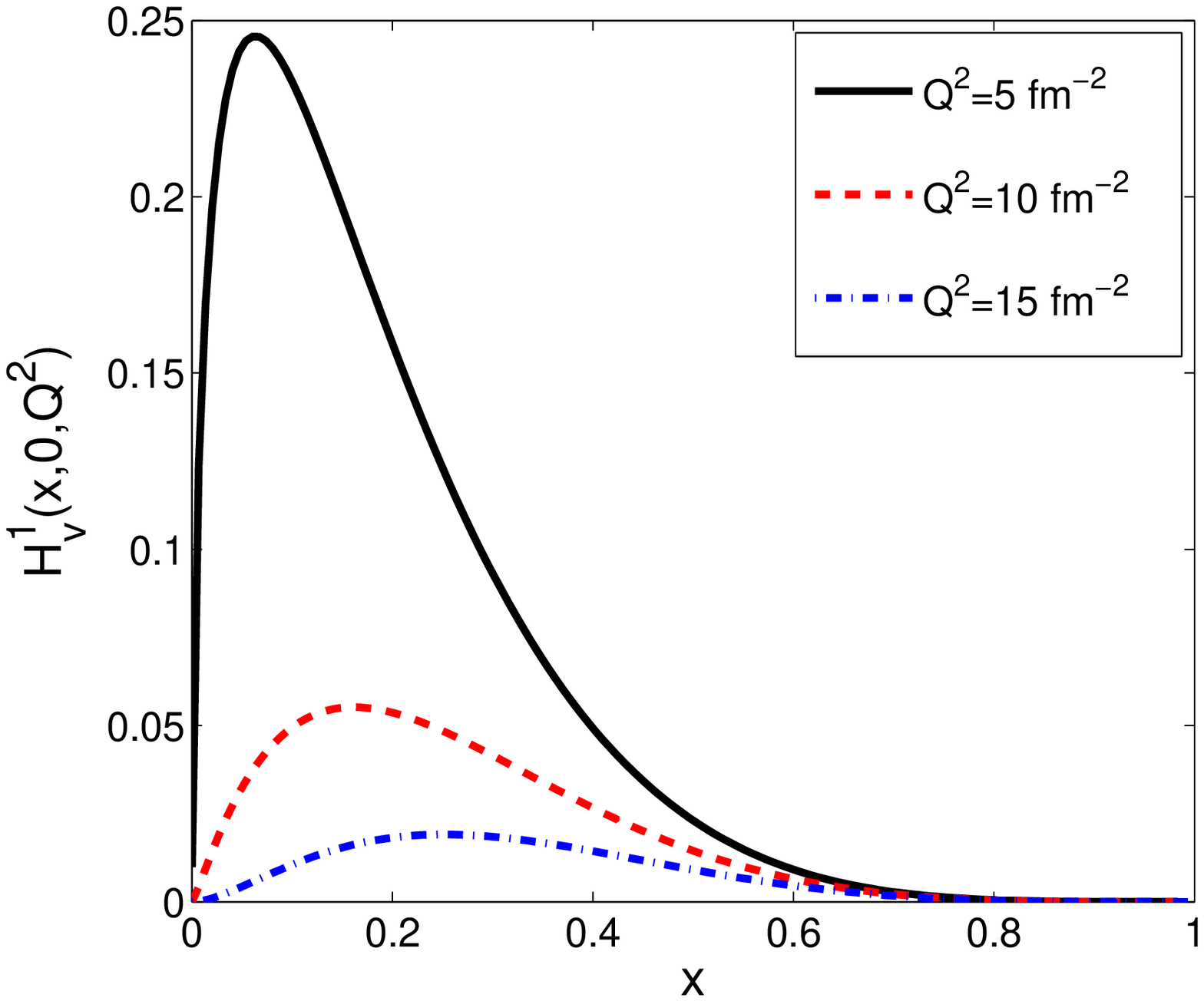}
\end{minipage}
\caption{\label{gpds_plot}(Color online) Plot of the deuteron GPD $H^1_v(x,0,Q^2)$ as a function of $x$ and $Q^2$. The other GPDs are $H^i_v(x,0,Q^2)=c_iH^1_v(x,0,Q^2),~ i=2,3$ where $c_2=1.714$, and $c_3=26.544$.  }
\end{figure*}
\begin{figure*}[htbp]
\begin{minipage}[c]{0.98\textwidth}
{(a)}\includegraphics[width=7.5cm,clip]{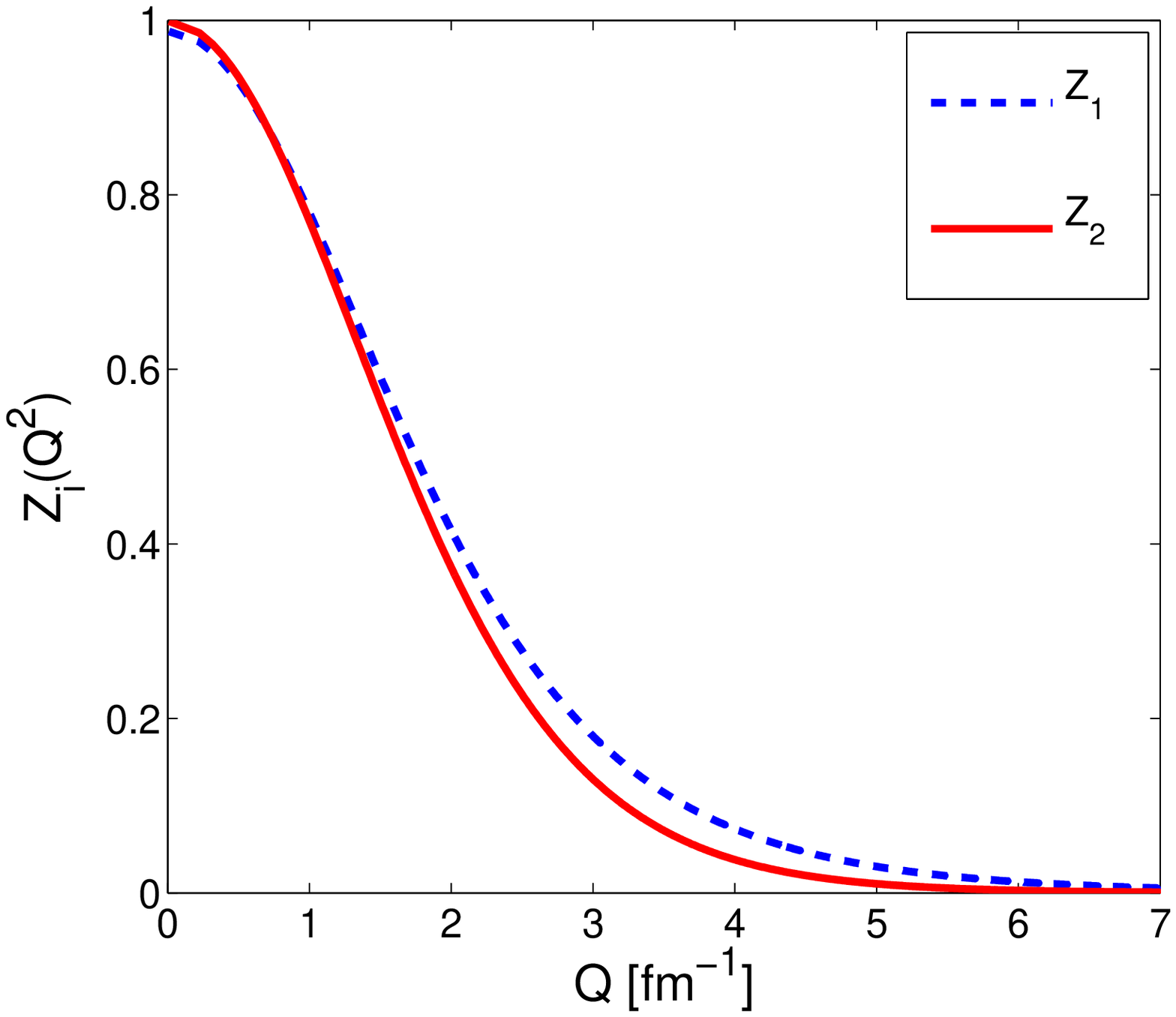}
{(b)}\includegraphics[width=7.5cm,clip]{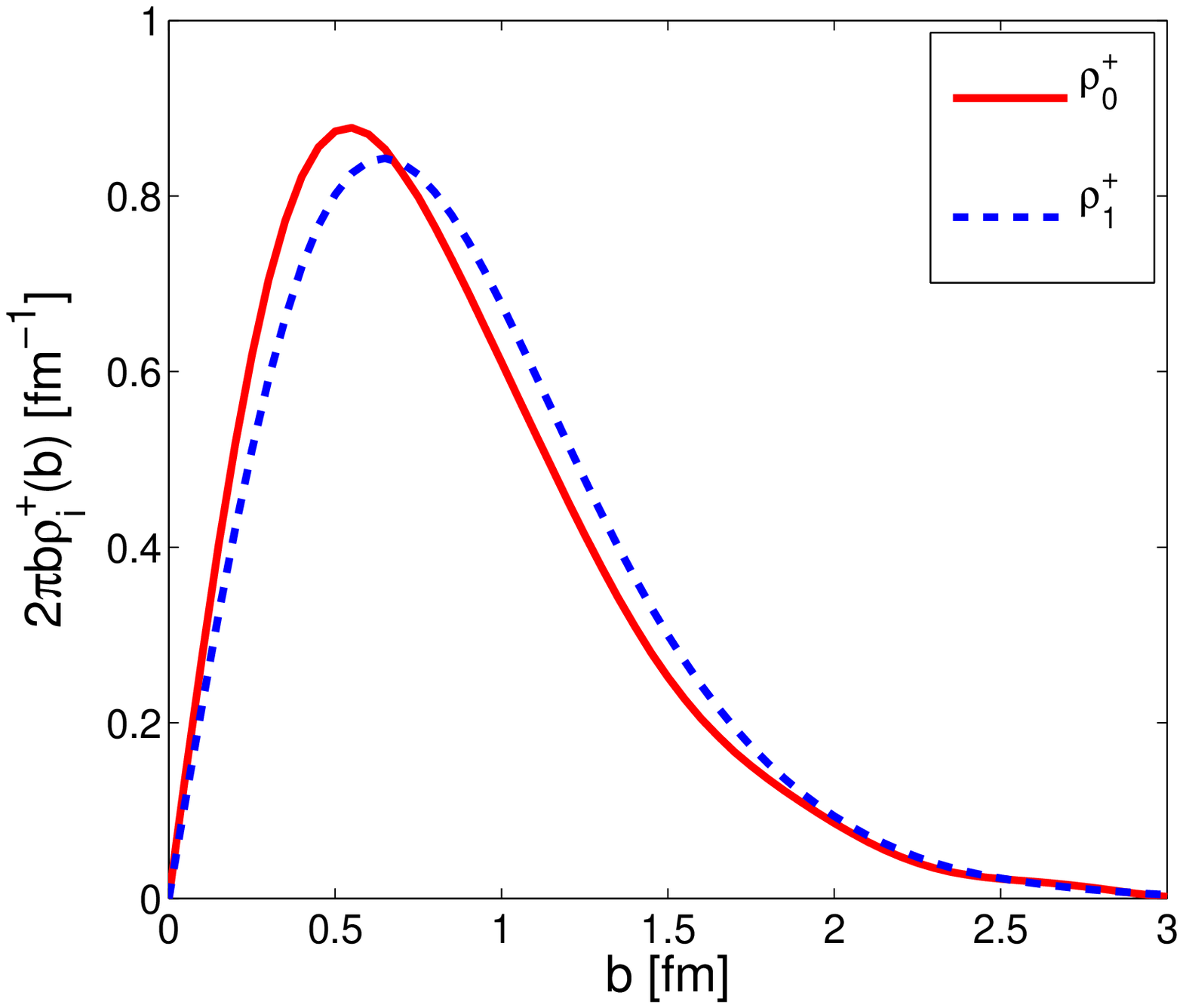}
\end{minipage}
\caption{\label{GFF}(Color online) Plots of (a) gravitational form factors for the deuteron, and (b) Longitudinal momentum densities in the deuteron.}
\end{figure*}
\section{Generalized parton distributions for the deuteron}\label{DGPDs}
Generalized parton distributions encode the information about the three dimensional spatial structure of the hadron
as well as the spin and orbital angular momentum of the constituents. GPDs are defined through the matrix element of the QCD non-local and non-diagonal operators on the light-front.
For the deuteron (spin-1 particle), there are five vector GPDs which are defined by~\cite{Berger:2001zb}
\be
&& \int \frac{p^+ dy^-}{2\pi} e^{ixp^+y^-} \left\langle p_2,\lambda_2 \right|  \bar q(-\frac{y}{2}) \gamma^+  
		q(\frac{y}{2})\left| p_1, \lambda_1 \right\rangle_{y^+=y_\perp = 0}
						\nonumber \\
&& = - 2 (\varepsilon_2^* \cdot \varepsilon_1) p^+ H_1
	- \left( \varepsilon_1^+ \, \varepsilon_2^* \cdot q 
		- {\varepsilon_2^+}^* \, \varepsilon_1 \cdot q \right) H_2 \nonumber\\
&&+ \ q\cdot \varepsilon_1 \, q\cdot \varepsilon_2^* \frac{p^+}{m_n^2} H_3
	- \left( \varepsilon_1^+ \, \varepsilon_2^* \cdot q 
		+ {\varepsilon_2^+}^* \, \varepsilon_1 \cdot q \right) H_4 \nonumber\\
&&	 +\ \left( \frac{m_n^2}{(p^+)^2} \varepsilon_1^+ \, {\varepsilon_2^+}^*
		+ \frac{1}{3} (\varepsilon_2^* \cdot \varepsilon_1) \right) 2 p^+ \, H_5 \,.
\ee
All the GPDs are functions of the three variables namely, the longitudinal momentum fraction $x$ carried
by the active quark, the square of the total
momentum transferred $t$ and the fraction of the longitudinal momentum transferred $\xi$, the so called
skewness. In symmetric frame the kinematical variables are $
p^{\mu}=\frac{(p+p')^{\mu}}{2}, ~q^{\mu}=p'^{\mu}-p^{\mu}, ~ \xi=-\frac{q^{+}}{2P^+},
$ and $t=q^2$. The first moments of GPDs are related to the form factors of the deuteron by 
\be
\int_0^1 dx \, H_v^i(x,\xi,t) = G_i(t) ,	\label{GPDs_FF}\quad 	&i&=1,2,3  \\
\int_0^1 dx \, H_v^i(x,\xi,t) = 0 ,	\quad	&i&=4,5	\,,
\ee
where the GPDs for the valence quark are defined as $H^i_v(x,0,t)\\=H^i(x,0,t)+H^i(-x,0,t)$. Using the integral representation of $V(z,Q^2)$ in Eq. (\ref{VInt}), one can rewrite the deuteron form factors in Eqs. (\ref{G1},\ref{Gi}) as
\be\label{FFs_int}
G_i(Q^2)=c_i\int_0^1 dx \int_0^\infty dz\frac{\kappa^{12} z^{11}}{12} \frac{x^{\frac{Q^2}{4\kappa^2}}}{(1-x)^2}e^{-\frac{\kappa^2z^2}{(1-x)}}.
\ee
Comparing the integrands in Eqs. (\ref{GPDs_FF}) and (\ref{FFs_int}), we extract
the GPDs for the deuteron in the following forms:
\be\label{GPDs}
H_v^i(x,0,Q^2)=c_i\int_0^\infty dz\frac{\kappa^{12} z^{11}}{12} \frac{x^{\frac{Q^2}{4\kappa^2}}}{(1-x)^2}e^{-\frac{\kappa^2z^2}{(1-x)}},
\ee
where $t=-Q^2$, and $c_1=1$, $c_2=1.714$, and $c_3=26.544$  \cite{Gutsche1}. Here we evaluate the deuteron GPDs when the skewness is zero ($\xi=0$). Since the GPD $H^4 (x,\xi,t)$ is odd in $\xi$, $H^4$ is zero for $\xi=0$ \cite{Berger:2001zb}. But the other GPDs are even in $\xi$. In Fig. \ref{gpds_plot}, we show the deuteron GPD $H^1_v(x,0,Q^2)$ as a function of $x$ and $Q^2$. The GPD has a peak at very low momentum fraction $x$, and also when the momentum transfer is low but the value decreases and goes to zero with increasing $x$ and $Q^2$. The other GPDs can be obtained by multiplying the constant $c_i$ with $H^1_v(x,0,Q^2)$.
\section{Longitudinal momentum densities in the transverse plane for the deuteron}\label{Ldensity}
Similar to the electromagnetic densities, we can identify the gravitomagnetic density ($p^+ ~momentum ~density$) in the transverse plane by taking the Fourier transform of the gravitational form factor \cite{selyugin,abidin08}. Thus, the longitudinal momentum ($p^+$) density for the deuteron can be defined as 
\be
\rho^+_\lambda(\vec{b}_\perp) &=& \int \frac{d^2\vec{q}_\perp}{(2\pi)^2} e^{-i\vec{q}_\perp\cdot\vec{b}}\mathcal{T}^+_{\lambda\lambda}(Q^2),
\ee
where the form factor $\mathcal{T}^+_{\lambda\lambda}(Q^2)$ is given by the matrix elements of the stress tensor
\be
&&\left<p^+, \frac{\vec{q}_\perp}{2},\lambda_2\Big| T^{++}(0) \Big|p^+,-\frac{\vec{q}_\perp}{2},\lambda_1\right>\nonumber\\
&\!=\!&
2(p^+)^2  \,  e^{i(\lambda_1-\lambda_2)\phi_q}
 \mathcal{T}^+_{\lambda_2\lambda_1}(Q^2).\label{Tmatrixelement}
\ee
For spin-1 particles, there are two independent helicity conserving form factors $\mathcal{T}^+_{11}$, and $\mathcal{T}^+_{00}$
which are given by \cite{abidin08,AC4}
\be
\mathcal{T}^+_{00}(Q^2)=Z_1(Q^2)&=&\int dz\mathcal{H}(Q,z) \partial_z\Phi_0(z) \partial_z\Phi_0(z)\,,\nonumber\\
\mathcal{T}^+_{11}(Q^2)=Z_2(Q^2)&=&\int dz \mathcal{H}(Q,z) \Phi_0(z) \Phi_0(z), \label{Z1Z2}
\ee
where the form of the profile function $\mathcal{H}(Q,z)$ in the soft-wall AdS/QCD model is \cite{abidin08,AC}
\be
\mathcal{H}(Q,z)=   \Gamma(4\eta+2) \, U(4\eta, -1, z^2),
\ee
where $U(a,b,w)$ is the $2{\rm nd}$ Kummer function and $\Phi_0(z)$ is the normalizable wave function, $\int dz \Phi_0^2=1$. We use the deuteron wave function (Eq.(\ref{wf})) in Eq.(\ref{Z1Z2}) to evaluate the gravitational form factors $Z_1(Q^2)$ and $Z_2(Q^2)$. Now, one can obtain the $p^+$ densities as
\be
\rho^+_1(\vec{b}_\perp)&=&\int_0^\infty \frac{dQ}{2\pi}Q J_0(bQ)Z_2(Q^2)\,,\nonumber\\
\rho^+_0(\vec{b}_\perp)&=&\int_0^\infty \frac{dQ}{2\pi}Q J_0(bQ)Z_1(Q^2).
\ee
We show the helicity conserving gravitational form factors $\mathcal{T}^+_{00}$ and $\mathcal{T}^+_{11}$ as a function of $Q^2$ in Fig. \ref{GFF}(a). One can notice that at $Q^2=0$, $\mathcal{T}^+_{\lambda,\lambda}$=1. Thus, it follows the momentum sum rule. The longitudinal momentum densities in the transverse plane are shown in Fig. \ref{GFF}(b). We observe that both the densities $\rho_1^+$ and $\rho_0^+$ are more or less same in the transverse plane. 

\section{\bf Summary}\label{concl}
In this work, we have studied the transverse charge densities in longitudinally and transversely polarized deuterons 
using the empirical formulae of the deuteron electromagnetic form factors evaluated in the framework of the soft-wall AdS/QCD. We found that the charge densities for longitudinally polarized deuterons are axially symmetric and exhibit the monopole pattern only. Transversely polarized deuterons show a monopole pattern together with the dipole and quadrupole structure 
in the charge distributions. The dipolar structure which appears only in the distribution for the transversely polarized deuteron state with spin projection $s_\perp=1$ causes a small overall shift in $\rho^d_{T \, 1}$ along the $y$-axis. The quadrupole structures stretch the charge distributions along the $y$-axis or $x$-axis but does not cause any shift from the {\it c.m.}. The monopole contributions to the charge densities are large compared to the dipole and quadrupole. Further, we have compared the AdS/QCD results with the consequences of phenomenological parameterization \cite{Carlson:2008zc}. Except for the mismatch in $\rho^d_{0}$ at the center and the overall small shift in $\rho^d_{T \, 1}$ which is opposite to phenomenological result, the comparison shows that the AdS/QCD predictions are in reasonable  agreement with the phenomenological parameterization. 

Then we have shown the GPDs for the deuteron obtained from the electromagnetic form factors using the integral representation form of the bulk to boundary propagator and the GPDs sum rule. We have also evaluated the helicity non-flip gravitational form factors and shown that they satisfy the physical condition at $Q^2=0$. Finally, the longitudinal momentum densities in the transverse plane have been investigated and we found that the longitudinal momentum densities are independent of the deuteron helicity. \\
\\
\\
This work is supported by new faculty startup funding by the Institute of Modern Physics, Chinese Academy of Sciences under the Grand No. Y632030YRC. 


\end{document}